\documentclass[
    reprint, 
    aps,
    preprintnumbers,
    twocolumn,
    prb,
    superscriptaddress
]{revtex4-2}

\usepackage[utf8]{inputenc}
\usepackage{booktabs}
\usepackage[multiple]{footmisc}
\usepackage{lipsum}
\usepackage{rotating}
\usepackage{perpage}
\usepackage{chronology}
\usepackage{amssymb}
\usepackage{amsbsy}
\usepackage{amsmath}
\usepackage{tikz}
\usepackage[T1]{fontenc}
\usepackage{etoolbox}
\usepackage{graphics}
\usepackage{abraces}
\usepackage{siunitx}
\usepackage{hyperref}
\usepackage{multirow}
\usepackage{mathtools}
\usepackage{bm}
\usepackage{url}
\usepackage{physics}
\usepackage{hyperref}
\usepackage{bbm}
\usepackage{color}
\usepackage{placeins}
\usepackage[normalem]{ulem}

\newcommand{\vk}{\vec{k}}
\newcommand{\vl}{\vec{l}}
\newcommand{\vQ}{\vec{Q}}
\newcommand{\up}{\uparrow}
\newcommand{\down}{\downarrow}
\newcommand{\kplusQ}{\vk+\vQ}
\newcommand{\kminusQ}{-\vk-\vQ}
\newcommand{\im}{\mathrm{i}}

\newcommand{\ddt}{\frac{\mathrm{d}}{\mathrm{d}t}}
\newcommand{\dgamma}{\mathrm{d}\gamma}

\newcommand{\mM}{\mathcal{M}}
\newcommand{\mN}{\mathcal{N}}

\newcommand{\greens}[1]{\mathcal{G}_\text{#1} (\omega)}
\newcommand{\green}[1]{\mathcal{G}_\text{#1}}
\newcommand{\spectral}[1]{\mathcal{A}_\text{#1}  (\omega)}

\newcommand{\bs}{\begin{subequations}}
\newcommand{\es}{\end{subequations}}
\newcommand{\be}{\begin{equation}}
\newcommand{\ee}{\end{equation}}

\def\hmath$#1${\texorpdfstring{{\rmfamily\textit{#1}}}{#1}}

\begin{document} 

\title{Collective excitations in competing phases in two and three dimensions}


\author{Joshua Alth\"user}\email{joshua.althueser@tu-dortmund.de}
\affiliation{Condensed Matter Theory, TU Dortmund University,
Otto-Hahn Stra\ss{}e 4, 44227 Dortmund, Germany}

\author{G\"otz S.~Uhrig}
\email{goetz.uhrig@tu-dortmund.de}
\affiliation{Condensed Matter Theory, TU Dortmund University,
Otto-Hahn Stra\ss{}e 4, 44227 Dortmund, Germany}

\date{\today}

\begin{abstract}
    We investigate the superconducting (SC), charge-density wave (CDW), and antiferromagnetic (AFM) phases in the 
		extended Hubbard model at zero temperature and half-filling. 
    We employ the iterated equations of motion approach to compute the two-particle Green's functions and their
		spectral densities. 
    This renders a comprehensive analysis of the behavior of collective excitations possible as the 
		model's parameters are tuned across phase transitions. 
    We identify the well-known amplitude (Higgs) and phase (Anderson-Bogoliubov) modes within the 
		superconducting phase and observe a similar excitation (cooperon) in the CDW phase which shifts towards 
		zero energy as the system approaches the phase transition to the SC phase. 
    In the CDW phase, close to the phase transition to the AFM phase, 
    we find a collective mode, an exciton, that does not change significantly 
    and another mode, a longitudinal magnon, 
    that emerges from the two-particle continuum as the system approaches the phase transition to the AFM phase. 
    It becomes identical with the former at the transition.
		In the AFM phase, their roles are reversed.
    Additionally, we find a transversal Goldstone magnon located at zero energy.
\end{abstract}

\maketitle


\section{Introduction}\label{sec:introduction}

The study of collective excitations is of great interest as it sheds light on the intricate dynamics of correlated electron systems, providing crucial insight into emergent material properties.
We are in particular interested in the behavior of the collective excitations in the vicinity of phase transitions.
Do they signal these transitions, for instance by softening? 
How does the competition of different phases manifest in the energies and weights of the collective excitations?
These questions set the context of the present article. 

On the theoretical side, we choose a paradigmatic, but simple model which displays competing phases.
The Hubbard model has been employed in a plethora of previous studies and competing phases are established in its extensions. 
Early studies proved the existence of eigenstates of the Hubbard Hamiltonian that exhibit off-diagonal long-range order, encouraging the model's usage for the description of high-temperature superconductivity \cite{yang89}.
Shortly afterward, an exact SO(4) symmetry was discovered, which induces a degeneracy of 
superconductivity (SC) and a charge-density wave (CDW) governed by an attractive on-site interaction \cite{yang90}.
This coexistence stems from the existence of a particle-hole transformation on bipartite lattices, 
that maps the attractive Hubbard model rigorously onto the repulsive one, 
exhibiting antiferromagnetism (AFM) \cite{Hirsch85}.
The order parameters of the SC and CDW phases in the attractive model map to different spin expectation values in the repulsive one \cite{zitko15,lieb89}.

Numerous studies investigated the phases and various quantities of the Hubbard model in equilibrium systems, including additional interactions with and without doping
\cite{Micnas88,Micnas88b,Micnas89,Dzierzawa92,Kostyrko92,Eriksson95,Staudt00,Onari04,Toschi05,Brackett16,Paki19,romer20,Sushchyev22}. 
Recent studies on the dynamics of the superconducting gap parameter examined quenches, yielding oscillations \cite{Volkov73,Yuzbashyan05,Yuzbashyan06,Barankov06,Cui19}, 
and driven systems motivated by the goal to induce superconductivity \cite{Nicoletti14,Krull14,Moor14,Casandruc15,patel16,sentef17,Buenemann17}.

In this paper, we restrict ourselves to the half-filled Hubbard model 
including an additional nearest neighbor, intersite interaction on the square and the simple cubic lattice at zero temperature.
Specifically, we investigate competing phases, i.e., the behavior of various collective modes close to phase transitions between these phases.
To this end, the case in two dimensions (2D) already displays a wide range of possible phases
including CDW, AFM, $s$- and $d_{x^2 - y^2}$-wave superconductivity as well as 
phase-separated states \cite{Micnas88b,Tsuchiura95,Su01,Su04,ha11,Huang13,Jiang22,Linner23}.
We extend the corresponding established phase diagrams to the simple cubic lattice in three dimensions
(3D) where we find a qualitatively similar phase diagran for CDW, AFM, $s$-wave SC phases, and
indications of phase separation.

First, we employ a mean-field approximation to the interaction terms to determine the phases.
Second, we use the iterated equations of motion approach (iEoM)
which has already seen success in the handling of interaction quenches \cite{uhrig09,hamerla13,hamerla14,bleicker18}
to compute the collective excitations.
The basic idea is to start in the Heisenberg picture from a suitable operator basis which is extended
upon commuting its operators with the Hamiltonian and including the appearing additional operators to the basis. 
Of course, a truncation is necessary for most practical applications, however, the approximation becomes better the more terms are included within the basis.
The applicability of this method was compared successfully to the results of the density matrix formalism \cite{Kalthoff17}.

Moreover, we demonstrate an explicit way to compute various Green's functions by this approach.
By extension, the corresponding spectral functions of the investigated systems can be computed and 
we discuss the signatures of collective excitations in the spectral functions. 
The most prominent examples in the SC phase are the well-known phase mode (Anderson-Bogoliubov) and 
the amplitude (Higgs) modes. The former occurs in neutral superfluids at zero energy and was found
in a large number of studies \cite{Bogoljubov1958,Anderson58,Brieskorn74,Schmid1975,simanek1975,schon76,Maiti2015,Sun2020,Fan22} which we cannot list exhaustively here.
We obtain this mode based on a microscopic description without long-range electromagnetic interactions.
The inclusion of this kind of interaction would shift this mode towards the plasma frequency 
\cite{Anderson58,schon76,Kulik1981} which is, however, beyond the scope of the present paper. 
The amplitude mode, on the other hand, is located at the lower edge of the quasiparticle continuum.
The corresponding energy is the energy required to break up a Cooper pair 
\cite{Schmid1975,Varma02,Cea14,Measson14,Tsuji15,Krull16,Mueller2019,Schwarz20}.
This Higgs mode is not charged so that it does not couple to the electromagnetic fields.

This paper is organized as follows:
In \autoref{sec:model} we introduce the model and its Hamiltonian as well as the employed mean-field theory for the ground state.
We give a brief overview of the iterated equations of motion approach and 
derive a rigorous relation to Green's functions in \autoref{sec:ieom}.
Next, we show and discuss results in \autoref{sec:results}.
Lastly, in \autoref{sec:conclusion}, we summarize the results, draw the conclusions, and provide
an outlook.

\section{Model and mean-field theory}\label{sec:model}

\subsection{Model}

In our study, we employ the extended Hubbard model at half-filling as it hosts
the relevant phases and thereby provides direct access to the rich excitation spectra therein.
Its Hamiltonian is given by
\begin{align}
\nonumber
        H = &-t \sum_{\langle i, j \rangle, \sigma} \left( c_{i\sigma}^\dagger c_{j\sigma} + \text{h.c.} \right) 
        + \mu \sum_{i,\sigma} n_{i\sigma} \\
        & + U \sum_{i} n_{i\uparrow} n_{i\downarrow} 
        + \frac{V}{2} \sum_{\langle i, j\rangle, \sigma} n_{i\sigma} n_{j\sigma},
				    \label{eqn:full_hamiltonian}
\end{align}
where $c_{i\sigma}^{(\dagger)}$ annihilates (creates) an electron with spin $\sigma$ on lattice site $i$ 
and $\langle i, j\rangle$ denotes the summation over nearest neighbor sites.
The parameters are the hopping amplitude $t$, the onsite interaction $U$, the intersite interaction $V$, and the chemical potential $\mu$.
Applying a Fourier transform passing to $k$ space yields the single-particle dispersion 
\begin{equation}
    \epsilon_0 (\vk) = -2t \sum_{\alpha=1}^D \cos(k_\alpha), \quad k_\alpha \in [-\pi, \pi),
\end{equation}
with the system's dimension $D$ and the dimensionless wave vector $\vk$ where we set the lattice constant to unity.

We investigate this model for various sets of parameters allowing us access to a variety of phases.
The model exhibits antiferromagnetism for positive $U$ and moderate values of $V$.
At larger $V$, the intersite repulsion takes over and favors a charge-ordered phase, i.e., a CDW.
In the $U<0$ region, the CDW occurs for any $V>0$.
Choosing $V=0$ leads to a coexistence of $s$-wave superconductivity and said CDW \cite{yang90}, while moderate
values of  $V<0$ yield an $s$-wave superconducting phase \cite{Micnas88b}.
In this article, we explore the AFM-CDW as well as the CDW-SC phase transitions.
For $U<0$ and $V<0$, signs of phase separation occur as one would expect.

\subsection{Static mean-field theory}

Next, we decouple the interaction terms according to Wick's theorem.
We define short hands for the operators
\bs
\begin{align}
    \label{eqn:operators}
        n_{k\sigma} &\coloneqq  c_{\vk\sigma}^\dagger c_{\vk\sigma}      &f_k     &\coloneqq  c_{-\vk\down} c_{\vk\up} \\
        g_{k\sigma} &\coloneqq  c_{\vk\sigma}^\dagger c_{\vk+\vQ\sigma}  &\eta_k  &\coloneqq  c_{-\vk-\vQ\down} c_{\vk\up}
\end{align}
\es
where $\vQ \coloneqq  (\pi, \pi)$ in 2D and $\vQ \coloneqq  (\pi, \pi,\pi)$ in 3D
defines the nesting vector for the CDW and AFM phases. We use the following abbreviations to write down the mean-field parameters
\begin{subequations}
    \begin{align}
        \label{eqn:delta_cdw}
        \Delta_\text{CDW} &= \left(\frac{U}{2N} - 
				\frac{zV}{N}\right) \sum_{\vk\sigma} \langle g_{\vk\sigma} \rangle \\
        \label{eqn:delta_afm}
        \Delta_\text{AFM} &= \frac{U}{2N} \sum_{\vk} \left( \langle g_{k\uparrow} \rangle - \langle g_{\vk\downarrow} \rangle \right), \\
        \Delta_\text{SC} &= \frac{U}{N} \sum_{\vk} \langle f_{\vk} \rangle, \\
        \Delta_\eta &= \frac{U}{N} \sum_{\vk} \langle \eta_{\vk} \rangle, \\
        \Delta_n &= \frac{V}{N} \sum_{\vk,\sigma} \sum_{\alpha=1}^D \cos k_\alpha \langle n_{\vk\sigma} \rangle,
    \end{align}
\end{subequations}
where $z$ denotes the coordination number of the lattice.
The last parameter, $\Delta_n$, renormalizes the hopping term according to 
\begin{equation}
    \epsilon( \vk ) \coloneqq  -(2 + \Delta_n) \sum_{\alpha=1}^D \cos(k_\alpha)\,.
\end{equation}
In total, we obtain the mean-field Hamiltonian in spinor representation as
\begin{equation}
    \label{eqn:mf_hamiltonian}
    H_\text{MF} = \sum_{\vk} \Psi^\dagger (\vk) h(\vk) \Psi (\vk),
\end{equation}
with the spinors
\begin{equation}
    \Psi^\dagger (\vk) \coloneqq  \left( c_{\vk\up}^\dagger, c_{\kplusQ\up}^\dagger, c_{-\vk\down}, c_{\kminusQ\down} \right)
\end{equation}
and the matrix
\begin{equation}
    h(\vk) \coloneqq  \begin{pmatrix}
        \epsilon (\vk) & \Delta_-^* & \Delta_\text{SC} & \Delta_\eta \\
        \Delta_- & \epsilon (\vk + \vQ) & \Delta_\eta & \Delta_\text{SC} \\
        \Delta_\text{SC}^* & \Delta_\eta^* & - \epsilon (-\vk) & - \Delta_+ \\
        \Delta_\eta^* & \Delta_\text{SC}^* & - \Delta_+^* & - \epsilon (-\vk - \vQ)
        \end{pmatrix},
\end{equation}
where we defined $\Delta_\pm \coloneqq \Delta_\text{CDW} \pm \Delta_\text{AFM}$.

\begin{figure}
    \centering
    \includegraphics[width=.48\textwidth]{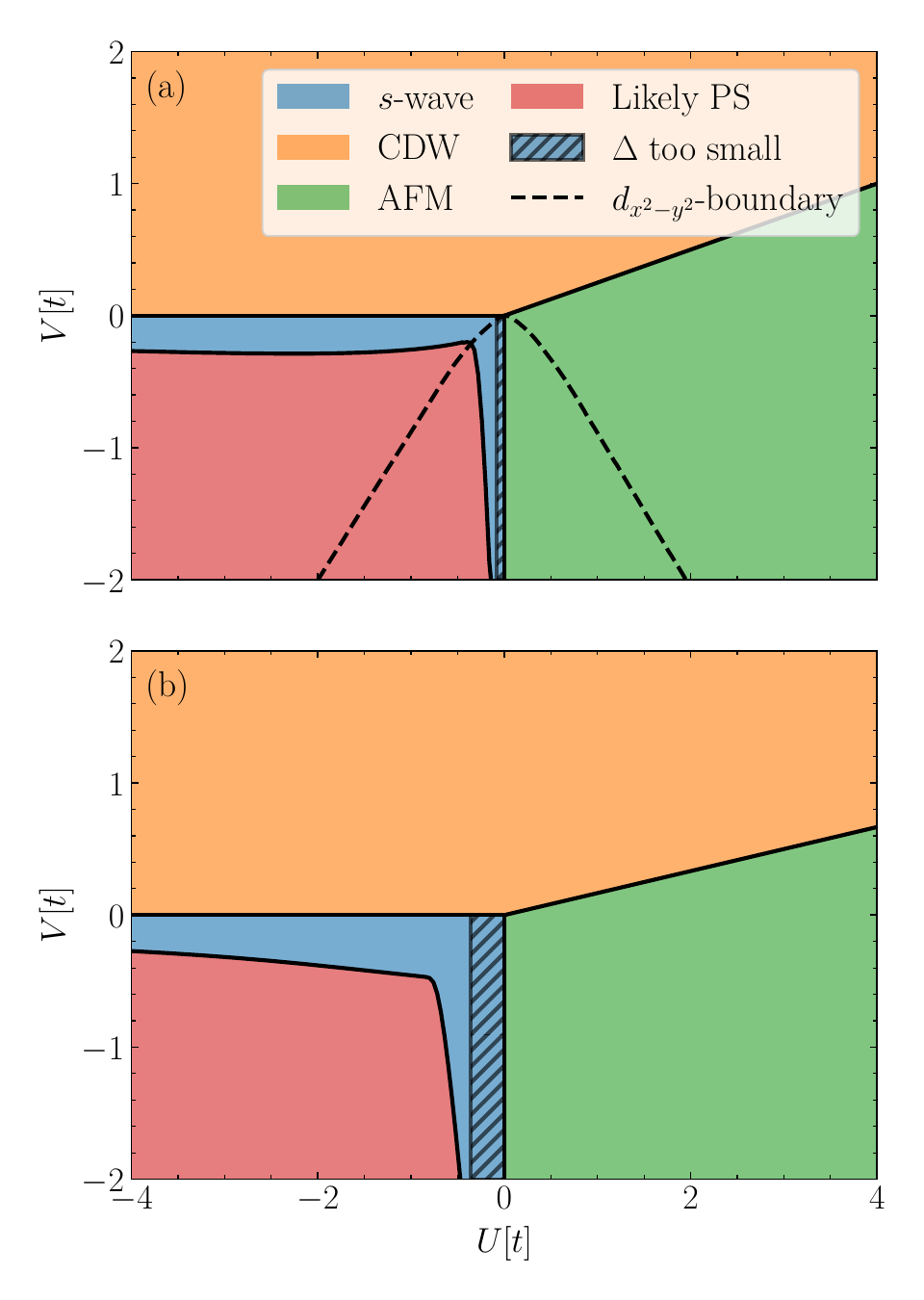}
    \caption{The phase diagram obtained for the extended Hubbard model \eqref{eqn:full_hamiltonian} on 
    (a) a square lattice and (b) a simple cubic lattice using a mean-field approximation at $T=0$.
    The CDW-AFM boundary lies at $U = zV$, where $z$ is the coordination number. 
    For $U<0$ and $V=0$,  CDW and $s$-wave SC coexist.
    The boundary for the $d_{x^2 - y^2}$-wave SC (dashed line) for the square lattice is  taken from Ref.\
		\cite{Micnas88b} since it is not accessible within our formalism, see below.
    However, the red shading indicates the region in which the dynamical matrix $\mM$ in 
		Eq.\ \eqref{eqn:heisenberg}, has negative eigenvalues indicating the instability of the assumed phase.
    In view of the attractive interactions, we expect a separation of phases \cite{Linner23}.
    Lastly, the striped area shows the region that we cannot access numerically
		because the gap values are too small for any tractable discretization. 
    Yet an SC phase is expected.}
    \label{fig:phase_diagram}
\end{figure}

We observe that  the entire Hamiltonian and all expectation values in the considered phases only depend on
\begin{equation}
    \widehat{\gamma}(\vk) \coloneqq \frac{1}{D} \sum_{\alpha=1}^D \cos(k_\alpha),
\end{equation}
i.e., for any operator $\widehat{O}$ the following relation holds 
\begin{equation}
    \label{eqn:equal_expecs}
    \langle \widehat{O}_{\vk} \rangle = \langle \widehat{O}_{\vk'} \rangle \eqqcolon \langle \widehat{O}( \gamma ) \rangle,
\end{equation}
if $\gamma= \widehat{\gamma}(\vk) = \widehat{\gamma}(\vk')$.
While 2D systems  of about $100\times100$ lattice sites can still be solved by evaluating sums over  wave vectors,
computing solutions for large three-dimensional systems becomes impossible due to the $N=L^3$ scaling.
This issue can be resolved by using the aforementioned fact and replacing the wave vector sums 
by energy integrals using the density of states (DOS) for $\gamma$ defined by
\begin{equation}
    \rho(\gamma) \coloneqq  \frac{1}{N} \sum_{\vk} \delta \left(\gamma - \widehat{\gamma} (\vk) \right)\,.
\end{equation}
As an example, we consider 
\begin{equation}
    \Delta_\text{SC} = \frac{U}{N} \int_{-1}^{1} \dgamma \rho(\gamma) \langle f( \gamma ) \rangle\,.
\end{equation}
Henceforth, we use the exact DOS for the square and the simple cubic lattice provided 
in Ref.\ \cite{Hanisch97}. 
This allows us to access both 2D and 3D models on equal footing.
We can compute any phase as long as it does not introduce another kind of wave vector dependence.

We solve the mean-field equations self-consistently.
The $\gamma$ integrals are approximated numerically using a $\tanh$-$\sinh$ quadrature \cite{takahasi73} 
terminating once $|1 - \int \dgamma \rho(\gamma)| < 10^{-13}$ is achieved
and reusing the computed sampling points and weights.
This method excels at dealing with the singularities in the DOS 
requiring merely a few hundred function evaluations to achieve the desired accuracy.

This procedure yields the ground state phase diagram at $T=0$ shown in Fig.\ 
\ref{fig:phase_diagram}.
The CDW-AFM boundary is located at $U = zV$, where $z$ is the coordination number.
This holds for both the square lattice ($z=4$) and the simple cubic lattice ($z=6$) and can be seen by comparing the prefactors in Eqs.\ \eqref{eqn:delta_cdw} and \eqref{eqn:delta_afm}:
Crossing $U = zV$ changes which prefactor of the two order parameters is larger.

Note, that the $d_{x^2 - y^2}$-superconducting state, which has been confirmed for the square lattice 
\cite{Micnas88b,Huang13}, cannot be described by our method because it would require to 
deal with expectations values which do not depend only on $\widehat{\gamma}(\vk)$.


\section{Iterated equations of motion and Green's functions}
\label{sec:ieom}

So far we discussed the static mean-field equations that describe the ground state.
These results are used to compute the quantities, namely the expectation values, necessary for the iterated equations of motion approach \cite{uhrig09,hamerla13,hamerla14,bleicker18}.
On this basis, we can describe two-particle quantities such as the various collective excitations of the system.

We start with an operator set $\mathcal{B}$ that ideally is complete with respect to commutation with 
the Hamiltonian $H$, i.e., any commutator $[H, A]$ for any $A \in \mathcal{B}$ can be represented by linear combinations of operators in $\mathcal{B}$.
Then, we can express any time-dependent operator in this set by
\begin{equation}
    \label{eqn:time_dependent_operator}
    A(t) = \sum_n a_n(t) A_n,
\end{equation}
where the coefficients $c_j(t)$ capture the entire time dependence. 
Inserting this into the Heisenberg equation of motion and applying a sort of operator scalar product $(A_i|\cdot)$ 
on both sides of the equation yields
\bs
\begin{align}
        \ddt A(t) = \sum_n \ddt a_n(t) A_n &= \im \sum_n a_n(t) [H, A_n] \\
        \Rightarrow \sum_n \underbrace{(A_i | A_n)}_{\coloneqq \mN_{in}} \ddt a_n(t) &= 
				\im \sum_n \underbrace{(A_i | [H, A_n])}_{\coloneqq \mM_{in}} a_n(t) \\
        \Rightarrow \mN \ddt \vec{a}(t) &= \im \mM \vec{a}(t).
				    \label{eqn:heisenberg}
\end{align}
\es
where $\vec{a}:=(a_1, a_2, a_3,\ldots)^\top$.
The matrices $\mN$ and $\mM$ contain all the energetic and dynamic properties of the system.
The former is referred to as norm matrix while the latter is called the dynamic matrix.
The advantage is that one can now handle simple matrices rather than operators 
acting on an enormous Hilbert space.
The dimension of these matrices depends on the number of operators in the set $\mathcal{B}$.

In practice, however, generic Hamiltonians do not allow for
a complete operator set $\mathcal{B}$ to exist because the commutations usually introduce terms which are not 
yet in $\mathcal{B}$, e.g., because they are of higher order.
For instance, commuting a bilinear term with a quartic Hamiltonian yields quartic terms. 
Iterating the commutation yields hexatic ones and so on.
One can include such additional terms and generically the results will improve the more
terms are considered in $\mathcal{B}$. In practice, one has to restrict the operators
to a suitable class of terms, thereby truncating certain terms. Here, we will restrict ourselves
to bilinear operators to capture the leading effects of collective behavior.

Furthermore, we use the symplectic product as operator product
\begin{equation}
\label{eqn:scalar_product}
    (A | B) \coloneqq  \langle [A^\dagger, B] \rangle,
\end{equation}
where the expectation values are taken with respect to the assumed phase of the mean-field Hamiltonian 
\eqref{eqn:mf_hamiltonian}.
Note that this is not a proper scalar product since it is not positive semidefinite.
For example, let us assume that for some operator $A$
\bs
\begin{equation}
(A | A) = \langle [A^\dagger, A] \rangle > 0,
\end{equation}
then
\begin{equation}
(A^\dagger | A^\dagger) = \langle [A, A^\dagger] \rangle = - (A | A)  < 0.
\end{equation}
\es
This does not pose a serious issue in our calculations but must be kept in mind.
In particular, it implies that the norm matrix is \emph{not positive} as one might have expected.
We emphasize, that the dynamical matrix  $\mM$ is computed
by commuting with the \emph{full} Hubbard Hamiltonian \eqref{eqn:full_hamiltonian} 
enabling us to capture collective behavior 
rather than using only the mean-field Hamiltonian \eqref{eqn:mf_hamiltonian}
which captures the one-particle dynamics, but misses the two-body dynamics.

The dynamical matrix $\mM$ is positive semidefinite if the system is in thermal equilibrium.
We derive this statement in  App.\ \ref{sec:positive_M}.
We use this fact to further enhance our ground state phase diagram.
If $\mM$ has even a single negative eigenvalue for a combination of parameters, 
we know that the assumed phase does not represent a true ground state.
Physically, this means that there are ``excitations'' which in fact lower the energy. 
Thus, the assumed ground state is not the lowest state and therefore unstable.
This happens for certain values $U<0$ and $V<0$, see the red shading in Fig.\ \ref{fig:phase_diagram}.
For finite temperatures on the square lattice, a phase-separated state as well as the coexistence of a phase-separated state with $s$-wave superconductivity has been found in this region \cite{Linner23}.
The phase boundary that we propose here agrees qualitatively well with the one found in the aforementioned reference.
We expect a similar set of phases to be present on the simple cubic lattice. 

While the existence of a negative eigenvalue of $\mM$ proves that the assumed phase is
not in thermal equilibrium, the absence of negative eigenvalues does not imply absolute stability,
but only local stability with respect to the considered operators.
For example, we do not find any negative eigenvalues of $\mM$ in the $U>0$, $V<0$ region despite previous studies indicating $d_{x^2 - y^2}$-wave superconductivity \cite{Micnas88b,Huang13}.

In order to treat the square and the simple cubic lattice on equal footing, we exploit 
\eqref{eqn:equal_expecs} to define the operator set in $\gamma$ space
\begin{equation}
    \label{eqn:ieom_basis_operator}
    A_\gamma \coloneqq \frac{1}{\sqrt{N}} \sum_{\vk} \delta (\gamma - \widehat{\gamma}( \vk )) A_{\vk},
\end{equation}
where $A_{\vk}$ represents each type of operator defined in Eqs.~\eqref{eqn:operators} or
their Hermitian conjugates if the operators are not Hermitian themselves.
Operators with different indices are treated separately, e.g., $n_{\gamma \up}$ and $n_{\gamma \down}$ 
are distinct operators in the studied set $\mathcal{B}$.
Additionally, we consider an analogous expression for the operator
\begin{equation}
    \tau_{k} \coloneqq c_{k \uparrow}^\dagger c_{k \downarrow}
\end{equation}
and its Hermitian conjugate.
This will allow us to describe transversal magnons, i.e., perpendicular 
deviations from the staggered magnetization in the AFM phase.

For the numerical treatment, we have to discretize $\gamma$.
Still, this allows us to obtain accurate results with drastically smaller matrices 
compared to an operator set based on a discretization of wave vectors.
This is crucial for dealing with three-dimensional lattices.
Practically, we choose $N_\gamma = 6000$ equally spaced sampling points.
A detailed explanation of the numerics is provided in the App.\ \ref{sec:numerical_ieom}.
Computationally, our method is limited by the number of terms included in our operator set 
as well as by the value of $N_\gamma$.
If the gap $\Delta_\text{tot}$ is of the same order of magnitude as the discretization $t \Delta \gamma$ the numerics become inaccurate.
This manifests specifically if a spectral function has strong features at minute energies, e.g., 
the phase mode in the superconducting phase.
In this case, the dynamical matrix $\mM$ spuriously displays negative eigenvalues that vanish if 
$N_\gamma$ is increased so that $\Delta \gamma$ is sufficiently smaller than $\Delta_\text{tot}$.
This happens if the parameters are chosen from the striped region in Fig.\ \ref{fig:phase_diagram}.
We observe that this issue only arises in the SC phase;
nevertheless, we expect similar inaccuracies in the AFM region for small values of $U$.

In the next step, we rearrange these operators in a way reminiscent of the $x$ and $p$ operators
of a harmonic oscillator
\begin{equation}
    X_i \coloneqq  A_i + A_i^\dagger,\quad P_i \coloneqq  A_i - A_i^\dagger.
\end{equation}
Then, the studied operator set is given by $\mathcal{B}_{XP} \coloneqq \{ X_i, \ldots, P_i, \ldots \}$.
If some operators would be 0, e.g., $P_i$ for any Hermitian $A_i$, or duplicate, e.g., certain $X_i$ for $g$-type terms where $g_{k\sigma} = g_{k+Q\sigma}^\dagger$, we omit these redundant operators.

Due to this particular choice of operators, the matrices occurring in \eqref{eqn:heisenberg} acquire a block structure
\begin{equation}
    \label{eqn:xp_set}
    \mM = \begin{pmatrix}
        \mathcal{K}_+ & \kappa \\ \kappa^\dagger & \mathcal{K}_-
    \end{pmatrix},\quad \mN = \begin{pmatrix}
        \Lambda_+ & \mathcal{L} \\ \mathcal{L}^\dagger & \Lambda_-
    \end{pmatrix},
\end{equation}
where the upper block refers to all $X_i$ operators and the lower block to all $P_i$ operators.
Furthermore, due to symmetry, the relations $\Im [\mathcal{K}_\pm] = \Re [\kappa] = 0$ and $\Re [\Lambda_\pm] = \Im [\mathcal{L}] = 0$ hold.
Since the mean-field Hamiltonian is real, all ensuing expectation values are real
and we can conclude both matrices have large empty blocks with zero elements: $\kappa=0$ and $\Lambda_\pm=0$.
This condition is fulfilled for all cases investigated in this article because the model displays
inversion symmetry and time-reversal symmetry or at least the combination of both in the AFM phase.

To study the collective excitations quantitatively, we investigate various retarded Green's functions 
\bs
\begin{equation}
    G_{AB}^\text{ret} (t) = - \im \langle [A(t), B] \rangle \Theta(t),
\end{equation}
where $\Theta(t)$ is the Heaviside function and their Fourier transforms
\begin{equation}
    \label{eqn:standard_gf}
    G_{AB}(z = \omega + \im 0^+) = -\im \int_0^{\infty} e^{\im z t} \langle [A(t), B] \rangle \mathrm{d}t.
\end{equation}
\es
The choice of the symplectic operator product enables us to write down the
matrix-valued Green's functions in terms of the introduced norm and dynamical matrix
\bs
\begin{align}
    \label{eqn:green_function}
    \mathcal{G}(z &= \omega + \im 0^+) = \mN \frac{1}{-z \mN - \mM} \mN \\
        &\coloneqq  -\mN \mathcal{R}(z) \mN,
\end{align}
where we introduced the resolvent
\begin{equation}
    \label{eqn:resolvent}
    \mathcal{R}(z) \coloneqq  \frac{1}{z \mN + \mM}.
\end{equation}
\es
The entries of this matrix $\mathcal{G}(z)$ are the Fourier-transformed Green's functions given in 
\eqref{eqn:standard_gf} with respect to the studied operators. For simplicity, we formulate the
relation between the Green's functions and the matrices for the operators $A_j$. Eventually, we 
use the block representation resulting for the operators $X_i$ and $P_i$.
The matrix element $(j,i)$ refers to
\bs
\begin{align}
    \mathcal{G}_{ji}(z = \omega +\im 0^+) &=  G_{A_j A_i^\dagger} (z) \\
        &= -\im \int_0^{\infty} \langle [A_j(t), A_i^\dagger(0)] \rangle e^{\im z t} \mathrm{d}t.
\end{align}
\es

Performing the  Fourier transformation yields
\bs
\begin{align}
    G_{A_j A_i^\dagger} (z) &= \im \int_0^\infty e^{\im z t} \langle [ A_i^\dagger(0), A_j(t) ] \rangle \mathrm{d}t \
		\\
        &= \im \int_0^\infty e^{\im z t} 
				\langle [ \sum_m a_{i,m}^*(0) A_m^\dagger, \sum_n a_{j,n}(t) A_n ] \rangle \mathrm{d}t 
		\\
        &= \im \int_0^\infty \sum_{mn} e^{\im z t} a_{i,m}^*(0) \underbrace{\langle [ A_m^\dagger, A_n ] \rangle}_{\equiv \mN_{mn}} a_{j,n}(t) \mathrm{d}t 
    \\
        &= \im \int_0^\infty e^{\im z t} \vec{a}_{i}^\dagger (0) \mN \vec{a}_{j}(t) \mathrm{d}t,
\end{align}
\es
where $\vec{a}_j(0)$ and $\vec{a}_i^\dagger (0)$ embody the initial conditions $A_j = \sum_n a_{j,n} A_n$ 
and $A_i^\dagger = \sum_n a_{i,n}^* A_n^\dagger$.
Assuming that $\mN$ is invertible we solve the differential equation \eqref{eqn:heisenberg} 
\begin{equation}
    \vec{a}_{j}(t) = \exp \left( \im \mN^{-1} \mM t \right) \vec{a}_{j}(0).
\end{equation}
Then, the final result reads
\bs
\begin{align}
    \label{eqn:green_derivation}
    G_{A_j A_i^\dagger} (z) &= \im  \int_0^\infty \vec{a}_{i}^\dagger (0) \mN 
		\exp \left( \im \left(\mN^{-1} \mM + z \right) t \right) \vec{a}_{j}(0) \mathrm{d}t 
		\\
        &= - \vec{a}_{i}^\dagger (0) \left[ \mN \frac{1}{\mN^{-1} \mM + z} \right] \vec{a}_{j}(0) 
				\\
        &= - \vec{a}_{i}^\dagger (0) \left[ 
				\mN \mathcal{R}(z) \mN \right] \vec{a}_{j}(0)
\end{align}
\es
with the resolvent $\mathcal{R}(z)$ from \eqref{eqn:resolvent}.
Obviously, the key task is computing the resolvent $\mathcal{R}$ efficiently.

To this end, we exploit the matrix structure in \eqref{eqn:xp_set} 
and assume that all matrix entries are real so that $\kappa_{ij} = \Lambda_{\pm,ij} = 0$ holds.
Then, with $r_X(z)\coloneqq\mathcal{R}|_{XX}$ denoting the upper left block of a matrix $\mathcal{R}$ which 
refers to the $X$-operators, we can calculate
\bs
\begin{align}
    r_X (z) & = \left. \frac{1}{\mM + z \mN} \right\vert_{XX} \\
        &= \left[ \frac{1}{\mM} - \frac{1}{\mM} z \mN \frac{1}{\mM} + 
				\frac{1}{\mM} z \mN \frac{1}{\mM} z \mN \frac{1}{\mM} - \cdots \right]_{XX}  \\
        &= \left. \frac{1}{\mM} \sum_{j=0}^\infty \left( -z \mN \frac{1}{\mM} \right)^j \right\vert_{XX} .
\end{align}
\es
We can safely omit every second term of the sum since we want to obtain the upper left block of the matrix only
and $\mN$ always swaps between the upper left and lower right block. Thus, we obtain
\bs
\begin{align}
    r_X (z) &= \left. \frac{1}{\mM} \sum_{j=0}^\infty z^{2j} \left( \mN \frac{1}{\mM} \right)^{2j} \right\vert_{XX} 
		\\
        &= \left. \frac{1}{\mM - z^2 \mN \mM^{-1} \mN} \right\vert_{XX} 
				\\
        &= \frac{1}{\mathcal{K}_+ - z^2 \mathcal{L} \mathcal{K}_-^{-1} \mathcal{L}^\dagger}.
\end{align}
\es
For brevity, we define $\check{N}_X \coloneqq \mathcal{L} \mathcal{K}_-^{-1} \mathcal{L}^\dagger$.
Since $\mathcal{M}$ is positive semidefinite and block diagonal its blocks $\mathcal{K}_\pm$ are positive semidefinite as well.
Therefore, $\check{N}_X$ is also positive semidefinite and Hermitian so that its square root can be defined.
Then, the resolvent $r_X(z)$ can be expressed by
\begin{equation}
    \label{eqn:rx}
    r_X (z) = -\check{N}_X^{-1/2} \frac{1}{z^2 - \check{N}_X^{-1/2} \mathcal{K}_+ \check{N}_X^{-1/2}} 
		\check{N}_X^{-1/2},
\end{equation}
where the matrix $\check{M}_X \coloneqq \check{N}_X^{-1/2} \mathcal{K}_+ \check{N}_X^{-1/2}$ is used; it is
positive semidefinite and Hermitian by the previous arguments.

If $\check{N}_X$ has a singular part, for instance, due to numerical inaccuracies or for particular operators, 
the inverse can be replaced by the Moore-Penrose pseudo-inverse. 
A possible way to compute this inverse is to diagonalize 
$\check{N}_X = \mathcal{V} \mathcal{D} \mathcal{V}^\dagger$ yielding a diagonal matrix $\mathcal{D}$ and a unitary transformation matrix $\mathcal{V}$. 
Then, the pseudo-inverse can be defined by
\begin{equation}
    \check{N}^{\mathrm{MP}} \coloneqq \mathcal{V} \mathcal{D}^{\mathrm{MP}} \mathcal{V}^\dagger
\end{equation}
with
\begin{equation}
    (\mathcal{D}^{\mathrm{MP}})_{ij} = \begin{cases}
        1/\mathcal{D}_{ij} & \mathcal{D}_{ij} \neq 0 \\ 0 & \text{otherwise}
    \end{cases}.
\end{equation}
Note that this definition recovers the standard inverse for invertible matrices.

The same calculation can be repeated by exchanging $X$ with $P$, yielding the relation for the lower right block.
The relevant matrices turn out to be 
\bs
\begin{align}
    \check{N}_P &\coloneqq \mathcal{L}^\dagger \mathcal{K}_+^{-1} \mathcal{L}
		\\
		\check{M}_P &\coloneqq \check{N}_P^{-1/2} \mathcal{K}_- \check{N}_P^{-1/2}\,.
\end{align}
\es
Summarizing, the Green's function is given by
\begin{subequations}
    \begin{align}
        G_{X_j,X_i}  (z) &= - \vec{x}_j^\dagger (0) \mathcal{L}^\dagger r_X (z) \mathcal{L} \vec{x}_i(0), 
				\\
        G_{P_j,P_i} (z) &= - \vec{p}_j^\dagger (0) \mathcal{L} r_P (z) \mathcal{L}^\dagger \vec{p}_i(0),
    \end{align}
\end{subequations}
where the vectors $x_j(0)$ and $x_i(0)$ describe the initial conditions for the operators
$X_j$ and $X_i$, respectively and $p_j(0)$ and $p_i(0)$ correspondingly for the operators
$P_j$ and $P_i$. The resolvent $r_X (z)$ is given in \eqref{eqn:rx} and 
\begin{equation}
    r_P (z) = -\check{N}_P^{-1/2} \frac{1}{z^2 - \check{M}_P} \check{N}_P^{-1/2} .
\end{equation}

The remaining task consists of finding the inverse $1/(z^2 - \check{M})$ for all $z$.
This, however, can be achieved efficiently using a Lanczos tridiagonalization and
a continued fraction expansion in terms of the Lanczos coefficients $a_i$ and $b_i$ 
\cite{PettiforRecursion,ViswanathRecursion}.
In the single-band case, the coefficients approach the limit
\begin{equation}
    \label{eqn:inf_lanczos}
    a_\infty = \frac{\omega_+ + \omega_-}{2}\text{  and  } b_\infty = \frac{\omega_+ - \omega_-}{4},
\end{equation}
where $\omega_\pm$ represents the upper and lower band edge, respectively \cite{PettiforRecursion}.
The continued fraction is truncated at some depth when the coefficients $a_i$ and $b_i$ are sufficiently
close to the limits \eqref{eqn:inf_lanczos}. Then, the truncated continued fraction can be terminated by the
square root terminator
\begin{equation}
\label{eqn:terminator}
    T(\omega) = \frac{1}{2b_\infty^2} 
		\left( \omega - a_\infty \mp \sqrt{(\omega - a_\infty)^2 - 4 b_\infty^2} \right),
\end{equation}
where the negative sign is to be chosen for $\omega - a_\infty > 2b_\infty$ and the positive sign for
$\omega - a_\infty < -2b_\infty$. For $|\omega - a_\infty| < 2b_\infty$, the square root in \eqref{eqn:terminator}
is replaced by $+\im\sqrt{4 b_\infty^2 - (\omega - a_\infty)^2}$.

Note that the coefficients start to deviate from the limits \eqref{eqn:inf_lanczos} as more and more Lanczos iterations are performed \cite{ViswanathRecursion}.
Therefore, we terminate the continued fraction with the aforementioned terminator when the pair of coefficients 
occurs in the fraction which deviates the least from $a_\infty$ and $b_\infty$.


\begin{figure*}
    \centering
    \includegraphics[width=.98\textwidth]{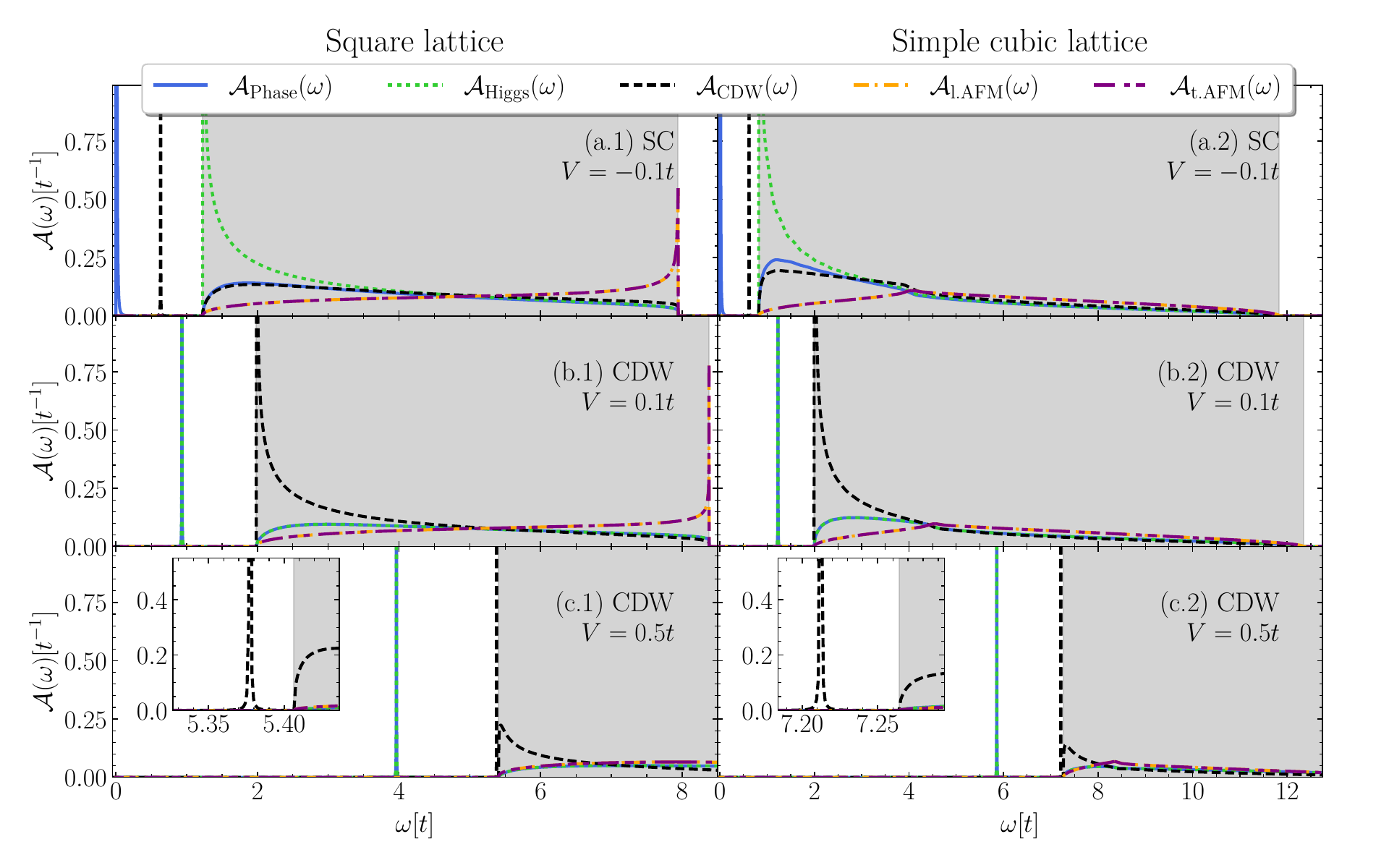}
    \caption{Spectral functions of the operators listed in Eq.~\eqref{eqn:resolvent_bases}.
    The left column shows the results for the square lattice and the right column for the simple cubic lattice.
    The left edge of each plot is at $-0.05t$  in order to improve the visibility of the phase peak at zero frequency
		($\hbar$ is set to unity). The gray area marks the two-particle continuum.
    The parameters are set to $U=-2.5t$ and $N_\gamma = 6000$, while $V$ is varied according to the legends.
    The panels show the spectral functions in the SC phase (a) and in the CDW phase, (b) and (c), respectively.}
    \label{fig:resolvent_overview_SC}
\end{figure*}

\begin{figure*}
    \centering
    \includegraphics[width=.98\textwidth]{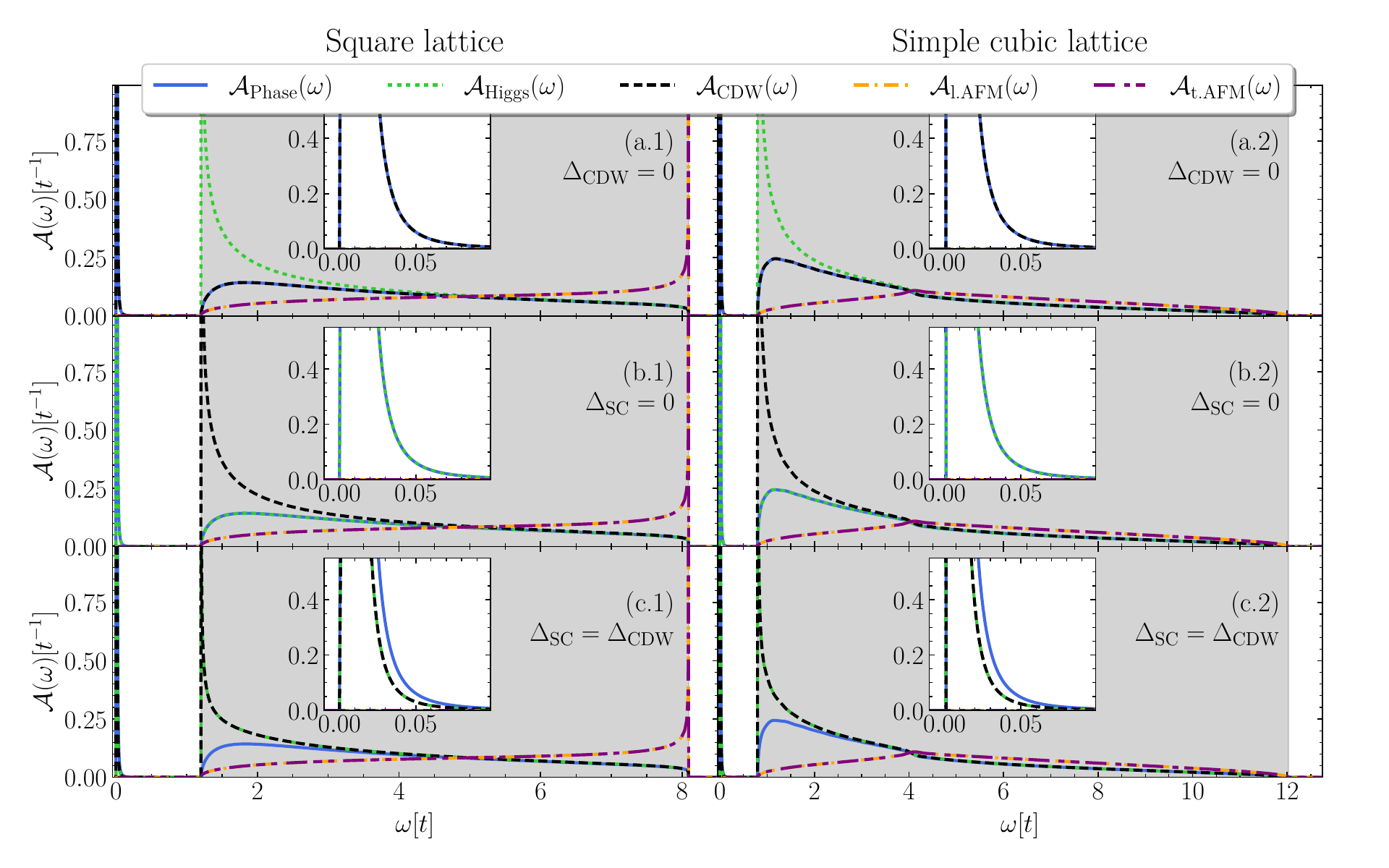}
    \caption{Same as Fig.\ \ref{fig:resolvent_overview_SC} except that  $V=0$, i.e., 
		SC and CDW order can coexist due to the SO(4) symmetry.
    Since the ratio of $\Delta_\text{SC}$ to $\Delta_\text{CDW}$ can be chosen arbitrarily, 
    we show in (a) the spectral functions for $\Delta_\text{CDW} = 0$ and in (b) for $\Delta_\text{SC} = 0$ while we distribute the gap equally in (c).}
    \label{fig:resolvent_overview_V0}
\end{figure*}

\begin{figure*}
    \centering
    \includegraphics[width=.98\textwidth]{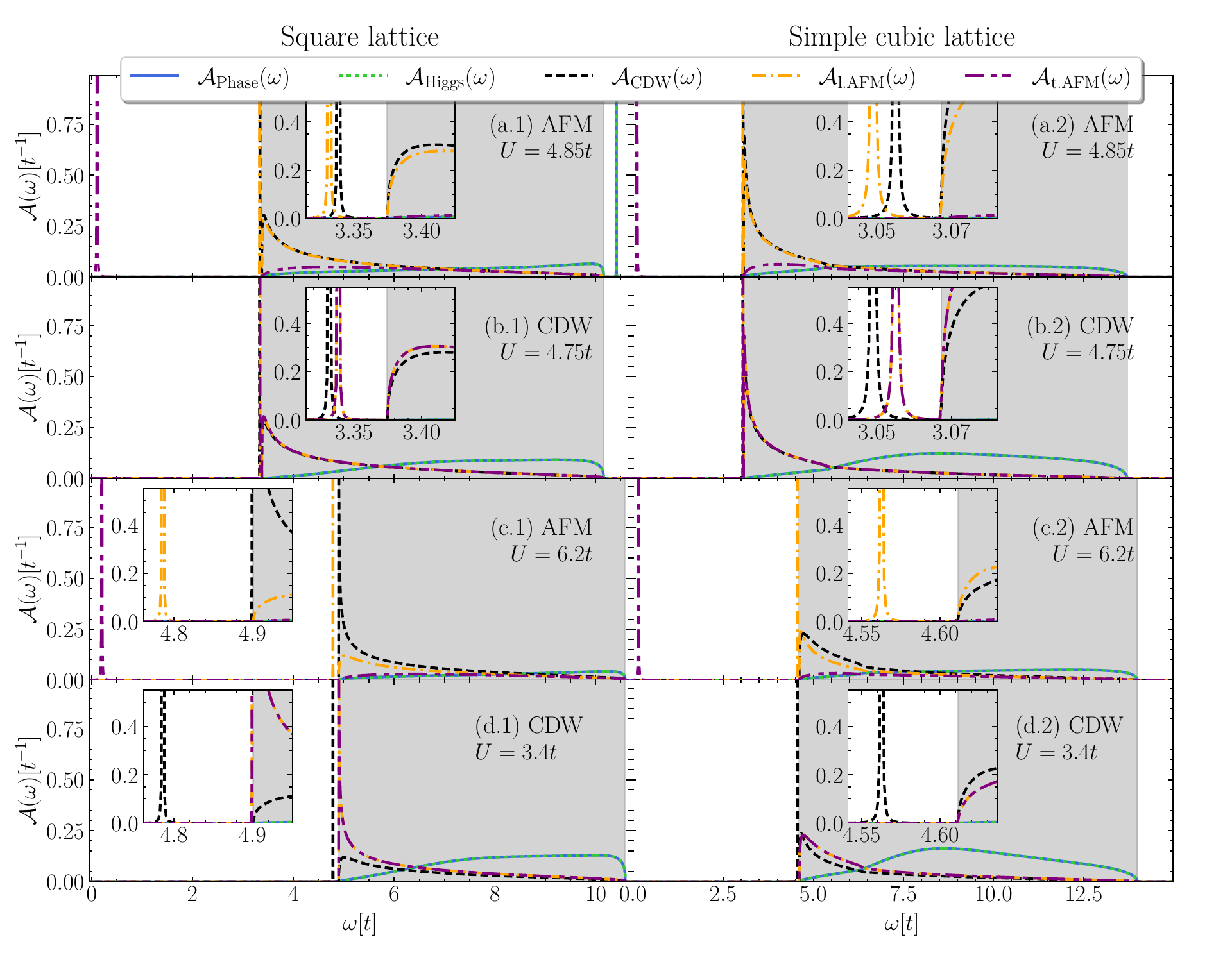}
    \caption{Same as Fig.\ \ref{fig:resolvent_overview_SC} except that we are focusing 
		on the AFM-CDW phase transition.
    For the square lattice, we set $V=1.2t$, and for the simple cubic lattice, $V=0.8t$. 
		This choice yields the phase transition at $U=4.8t$ for both lattices.
    The panels (a) and (c) show the spectral functions in the AFM phase and the 
		panels (b) and (d) in the CDW phase.}
    \label{fig:resolvent_overview_AFM}
\end{figure*}

\begin{figure*}
    \centering
    \includegraphics[width=.98\textwidth]{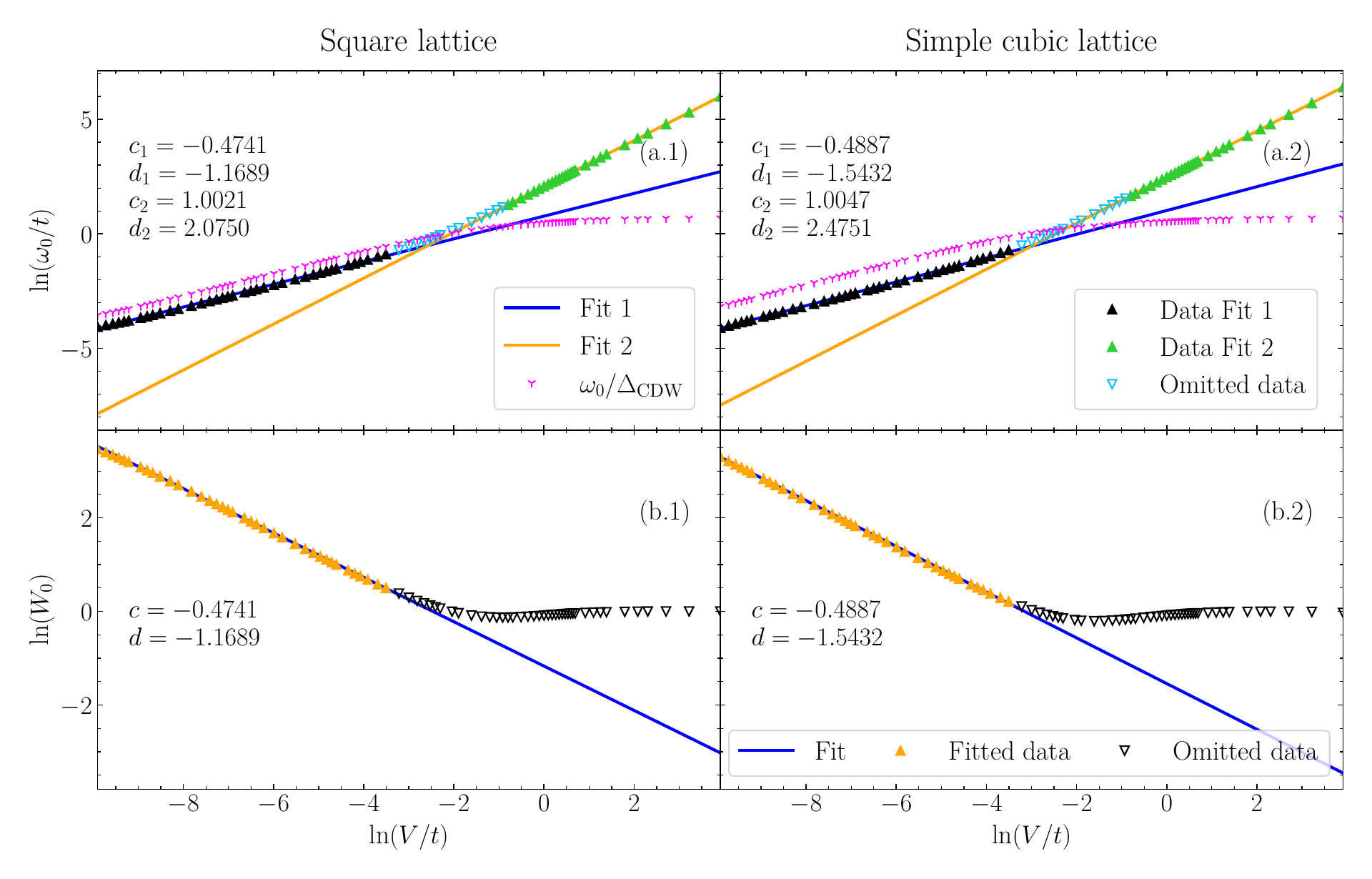}
    \caption{The upper panels (a) show double logarithmic plots of the positions $\omega_0$ of the peak in $\spectral{SC}$ in the CDW phase 
    while the lower panels (b) show their respective weights $W_0$.
    In the upper panels, the pink markers additionally show the peak position divided by the gap $\omega_0 / \Delta_\text{CDW}$. 
    The left column shows the results for the square lattice and the right column for the simple cubic lattice at $U=-2.5t$. 
    The fits are linear in the double logarithmic plot, i.e., $y(V) = e^d V^c$, and the parameters are indicated in the panels.}
    \label{fig:sc_in_cdw_behavior}
\end{figure*}

\begin{figure*}
    \centering
    \includegraphics[width=.98\textwidth]{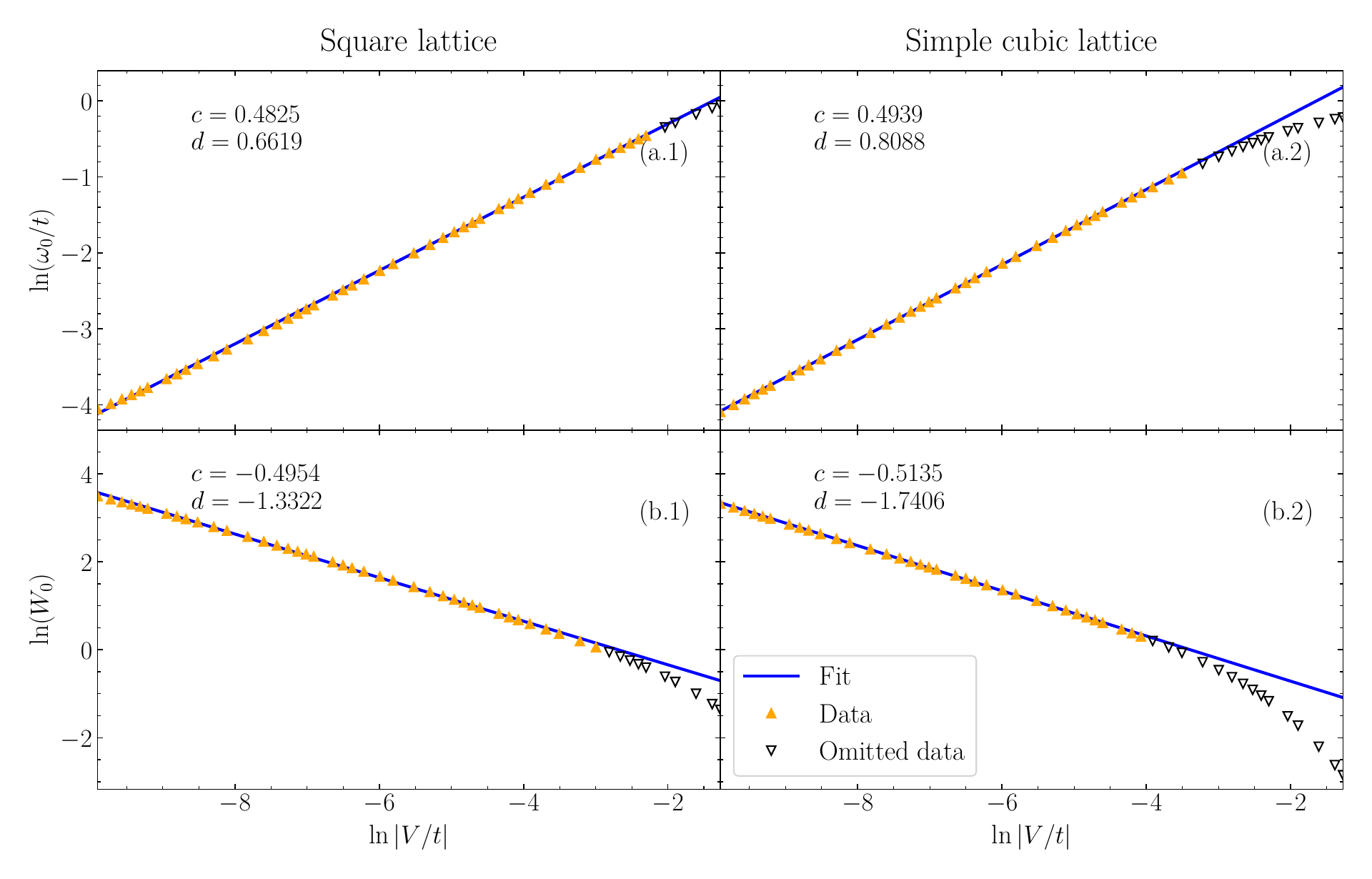}
    \caption{Same as in Fig.\ \ref{fig:sc_in_cdw_behavior}, but for the peak in $\spectral{CDW}$ 
		in the SC phase, i.e., for $V<0$.}
    \label{fig:cdw_in_sc_behavior}
\end{figure*}

\begin{figure*}
    \centering
    \includegraphics[width=.98\textwidth]{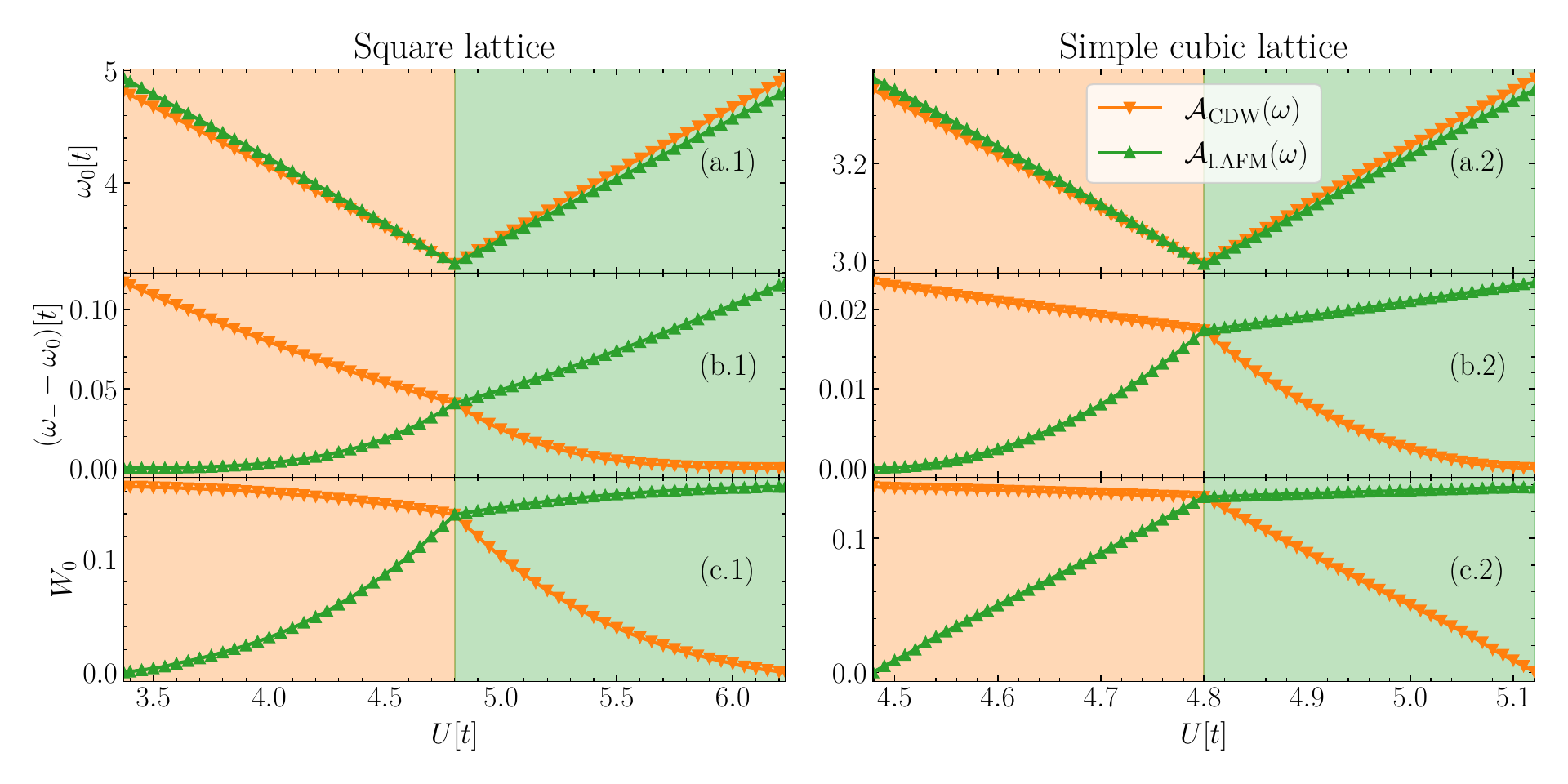}
    \caption{The upper panels (a) show plots of the positions $\omega_0$ of the peak in $\spectral{CDW}$ (orange lines) 
		and $\spectral{l.AFM}$ (green lines) close to the corresponding phase transition 
		at $U = 4.8t$.   Again, we choose $V=1.2t$ for the square lattice (left column) and $V=0.8t$ for the simple cubic lattice (right column), respectively.
    The middle panels (b) depict the peak positions relative to the lower edge of the two-particle continuum 
		$\omega_-$, while the bottom panels (c) show the peak weights.
    The shadings indicate the phase in which the system is. Green represents AFM while orange represents CDW.
    Note the difference in the scales for the two lattices.}
    \label{fig:afm_cdw_peaks_overview}
\end{figure*}

\begin{figure*}
    \centering
    \includegraphics[width=.98\textwidth]{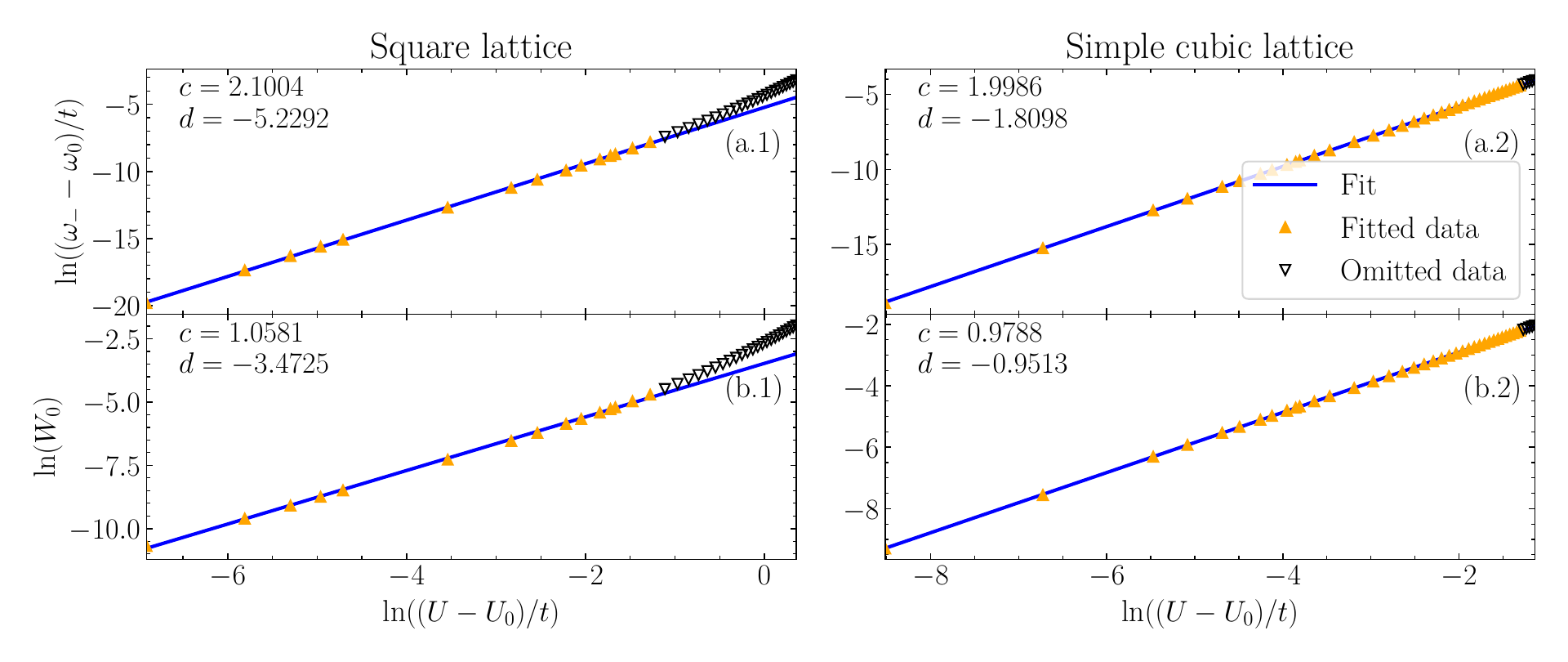}
    \caption{The upper panels (a) show double logarithmic plots of the positions $\omega_0$ 
		of the peak in $\spectral{l.AFM}$ relative to the lower edge of the two-particle continuum $\omega_-$.
    The lower panels (b) depict the respective weights $W_0$.
    Again, we choose $V=1.2t$ for the square lattice (left column) and 
		$V=0.8t$ for the simple cubic lattice (right column), respectively.
    The interaction $U_0$ denotes the highest value of $U$ for which 
		there is no peak below the two-particle continuum, i.e., for $U> U_0$ a peak occurs.
    For the square lattice, we find $U_0 \approx 3.372t$ and for the simple cubic lattice $U_0 \approx 4.479t$.
    The parameter range depicted here places the system within the CDW phase.
    The fits are of the same kind as in Fig.\ \ref{fig:sc_in_cdw_behavior}.}
    \label{fig:afm_cdw_peaks_details}
\end{figure*}



\section{Results}
\label{sec:results}

We investigate four different diagonal Green's functions $G_{AA^\dagger}(\omega + \im 0^+)$ with
\begin{subequations}
    \label{eqn:resolvent_bases}
    \begin{align}
        A_\text{Higgs} &= \frac{1}{\sqrt{N}} \sum_{\vk} \left( f_{\vk} + f_{\vk}^\dagger \right)  
            = \int_{-1}^1 \dgamma \left( f_{\gamma} + f_{\gamma}^\dagger \right) ,\\
        A_\text{Phase} &= \frac{i}{\sqrt{N}} \sum_{\vk} \left( f_{\vk} - f_{\vk}^\dagger \right) = 
				\int_{-1}^1 \dgamma \left( f_{\gamma} - f_{\gamma}^\dagger \right) ,\\
        A_\text{CDW}   &= \frac{1}{\sqrt{N}} \sum_{\vk} \left( g_{\vk \up} + g_{\vk \down} \right) = 
				\int_{-1}^1 \dgamma \left( g_{\gamma \up} + g_{\gamma \down} \right) ,\\
        A_\text{l.AFM}   &= \frac{1}{\sqrt{N}} \sum_{\vk} \left( g_{\vk \up} - g_{\vk \down} \right) = 
				\int_{-1}^1 \dgamma \left( g_{\gamma \up} - g_{\gamma \down} \right) , \\
        A_\text{t.AFM}   &= \frac{1}{\sqrt{N}} \sum_{\vk} \left( \tau_{\vk} + \tau_{\vk}^\dagger \right) = 
				\int_{-1}^1 \dgamma \left( \tau_{\gamma \up} + \tau_{\gamma \down} \right) ,
    \end{align}
\end{subequations}
where $N$ is the number of lattice sites in the system. 
Each of these operators generates a different kind of collective mode.
The first operator excites the amplitude mode of the $s$-wave superconducting state
while the second one excites the phase mode \cite{Fan22}.
The remaining two operators induce the collective behavior of the CDW and AFM order, respectively.
This means these Green's functions are the susceptibilities towards alternating local potential (CDW) or
alternating magnetic field (AFM).
Below, we refer to the Green's functions of these operators by
$\mathcal{G}_{\alpha}(\omega) = G_{A_\alpha A_\alpha^\dagger}(\omega)$ 
with $\alpha \in \{ \text{Higgs}, \text{Phase}, \text{CDW}, \text{l.AFM}, \text{t.AFM} \}$.
Furthermore, we use $\text{SC}$ when the Higgs and phase mode yield the same spectral function and
$\text{AFM}$ when longitudinal and transversal magnons yield the same spectral function.

The operators in \eqref{eqn:resolvent_bases} are Hermitian and of bosonic character. 
This means that all spectral functions are antisymmetric \cite{rickayzen80}.
For this reason, we only show the part  $\omega \geq 0$ in our plots. 
Additionally, we add a small positive imaginary part of $10^{-5}t$ to $\omega$ 
in order to plot the $\delta$ peaks as Lorentzians.
We analyze the effect of different numbers of sampling points $N_\gamma$ in App.\ \ref{sec:finite_size}.


\subsection{Classification of the spectral functions}

Let us describe the spectral functions 
$\mathcal{A}_\alpha (\omega) = - (1/\pi) \Im [\mathcal{G}_\alpha (\omega + \im 0^+)]$ of
the operators above in the various phases.
Fig.\ \ref{fig:resolvent_overview_SC} shows them at $U = -2.5t$ in the SC ($V=-0.1t$) and  in the 
CDW ($V=0.1t$ and $V=0.5t$) phase.

There are a variety of different features. 
The results for the lattices in 2D and 3D are very similar and differ only quantitatively.
In the SC phase, see Fig.\ \ref{fig:resolvent_overview_SC}(a), 
there is a sharp peak located at $\omega=0$ in $\spectral{Phase}$ and a 
singularity in $\spectral{Higgs}$ located at $\omega=2\Delta_\text{SC}$.
We identify them with the well-known Anderson-Bogoliubov mode and Higgs mode, respectively, in superconductors 
\cite{Bogoljubov1958,Anderson58,Brieskorn74,Schmid1975,simanek1975,schon76,Kulik1981,Maiti2015,Sun2020,Fan22,Schmid1975,Varma02,Cea14,Measson14,Tsuji15,Krull16,Mueller2019,Schwarz20}.
The former must not have any weight because of the antisymmetry of the spectral function.
The peak itself is located at zero energy because the model \eqref{eqn:full_hamiltonian} 
describes a neutral superfluid without coupling of the phase of the order parameter to the electromagnetic fields.
Therefore, this phase may be chosen arbitrarily, and changing it requires no energy 
which ultimately induces a Goldstone mode to exist \cite{Goldstone1961,Anderson63}.

To characterize the occurring peaks that lie outside the two-particle continuum, we inspect the real part of the Green's function.
We plot it in the vicinity of the peaks with a double logarithmic scale and fit 
the result linearly $y = ax + b$ to obtain its power-law behavior.
The real part is related to the imaginary part via the Kramers-Kronig relations.
Specifically, if the imaginary part is a $\delta$ distribution, the real part will be proportional to $1/\omega$, i.e., $a=-1$, and the peak has the weight $W_0 = \exp(b)$.
Furthermore, an exponent of $a=-2$ corresponds to the 
derivative of the $\delta$ distribution, i.e., $\delta'(\omega)$.
The fits themselves are given in App.\ \ref{sec:fit_greens_functions}.

This analysis yields that the real part of $\greens{Phase}$ near $\omega=0$ behaves like
$1/\omega^2$ showing that the peak in $\spectral{Phase}$ is the derivative of a $\delta$ distribution.
The weight of such a peak is given by the integral $\int \delta'(\omega) \mathrm{d}\omega = 0$ and thus
vanishes in accordance with expectations.

The singularity in $\spectral{Higgs}$ behaves like $1/\sqrt{\omega - 2 \Delta_\text{SC}}$.
Previous studies on the dynamics of the order parameter found dephasing oscillations after a quench.
These oscillations have the frequency $\Omega = 2 \Delta_\text{SC}$ and fall off like
$1/\sqrt{t}$ \cite{Volkov73,Kulik1981,Yuzbashyan06}.
An inverse Fourier transform of $\spectral{Higgs}$ yields precisely this behavior, 
thereby corroborating our results.

In the CDW phase, see panels (b) and (c) of Fig.\ \ref{fig:resolvent_overview_SC},
both spectral functions $\spectral{Higgs}$ and $\spectral{Phase}$ become identical and display a $\delta$ peak below the two-particle continuum.
In this case, both modes describe a cooperon, i.e., a bound state of two electrons or two holes.
Following the Landau theory for continuous phase transitions, it is to be expected 
that both excitations show the same spectral behavior. 
If the free energy is expanded in powers of the order parameter 
\begin{equation}
    f(\Delta_\text{SC}) \approx r |\Delta_\text{SC}|^2 + \frac{u}{2} |\Delta_\text{SC}|^4,
\end{equation}
with  $u > 0$. For $r<0$, the system is in the SC phase and $\Delta_\text{SC}$ is finite.
Then it makes a difference if one excites \emph{along} $\Delta_\text{SC}$ in the complex plane (Higgs mode)
or \emph{perpendicular} to it (phase mode). For $r>0$, however, the order vanishes, $\Delta_\text{SC}=0$,
and no distinction between Higgs and phase mode can be made \cite{Coleman2015}. 
Both possible excitation processes have to overcome the same energy. 

In the SC phase, $\spectral{CDW}$ features a $\delta$ peak below the continuum corresponding to the finite energy necessary to excite the electronic density modulation of CDW type.
This mode corresponds to the creation of an exciton since it results from a bilinear operator with
creation and annihilation fermionic operator so that it represents a bound electron-hole pair.
For moderate values of $V>0$, i.e., in the CDW phase, see panel (b) at $V=0.1t$, the CDW spectral function 
shows a singularity at $\omega=2\Delta_\text{CDW}$.
This mode moves out of the two-particle continuum upon increasing $V$ 
and becomes a proper $\delta$ peak, see panel (c) of Fig.\ \ref{fig:resolvent_overview_SC} at $V=0.5t$.

The spectral functions $\spectral{t.AFM}$ and $\spectral{l.AFM}$ are identical and restricted to 
the continuum for both phases and both lattices.
This can be understood by the same argument used for the identity of 
$\spectral{Phase}$ and $\spectral{Higgs}$ outside of the SC phase:
If no AFM order is present, there is no distinction between a longitudinal and a transversal excitation. 
The same amount of energy is required to create any such excitation which is a magnon
or paramagnon if there is no long-range AFM order.
For the square lattice, there is a singularity located at the upper edge of the two-particle continuum.
But this does not occur in the 3D case. Including lifetime effects, sharp features at the upper
edge of the continuum are expected to be smeared out anyway.


At $V=0$, the SC and the CDW order can coexist due to the SO(4) symmetry. 
In this phase, the mean-field single-particle energies are given by
\begin{equation}
    E_{\pm} = \pm \sqrt{\epsilon^2 + |\Delta_\text{CDW}|^2 + |\Delta_\text{SC}|^2}.
\end{equation}
This means that the proper order parameter is $\Delta_\text{tot} 
\coloneqq \sqrt{|\Delta_\text{CDW}|^2 + |\Delta_\text{SC}|^2}$.
Here, $\Delta_\text{CDW}$ and $\Delta_\text{SC}$ can change arbitrarily without affecting the system's energy as long as $\Delta_\text{tot}$ remains constant.
This degeneracy of both phases is rigorously exact due to the SO(4) symmetry \cite{yang90}. 

In Fig.\ \ref{fig:resolvent_overview_V0}, we show the spectral functions for $U=-2.5t$ and $V=0$.
The different panels correspond to different choices for $\Delta_\text{CDW}$ and $\Delta_\text{SC}$.
First, in panel (a), we set $\Delta_\text{CDW} = 0$.
This results in a purely superconducting phase with the same features as in 
Fig.\ \ref{fig:resolvent_overview_SC}(a).
The peak in $\spectral{CDW}$ moves to $\omega=0$ because it does not cost energy to enhance the CDW order parameter
if it is compensated by a reduction of the SC order parameter. 
In fact, $\spectral{CDW}$ and $\spectral{Phase}$ are identical since they are linked by the SO(4) symmetry.

Second, panel (b) shows the spectral functions for $\Delta_\text{SC} = 0$, i.e., in a purely charge-ordered phase.
Similar to before, there is not much change compared to Fig.\ \ref{fig:resolvent_overview_SC}(b).
The spectral functions $\spectral{Higgs}$ and $\spectral{Phase}$ are still identical since no
SC order is present. Both exhibit a sharp peak at $\omega=0$ proportional to $\delta'(\omega)$
which reflects that an SC order can be introduced without energy cost if compensated by a reduction of the CDW order.

Last, we choose $\Delta_\text{CDW} = \Delta_\text{SC}$ in panel (c).
For this specific ratio of $\Delta_\text{CDW}$ and $\Delta_\text{SC}$, 
$\spectral{Higgs}$ and $\spectral{CDW}$ are identical.
Choosing a different ratio yields qualitatively the same results, but the peaks differ in magnitude.
Both spectral functions have a peak at $\omega = 0$ proportional to $\delta'(\omega)$
and a singularity at the lower edge of the two-particle continuum $\omega = 2\Delta_\text{tot}$.
These singularities behave like $1/\sqrt{\omega - 2 \Delta_\text{tot}}$.
Varying the ratio of the two gap parameters allows one to
switch between $\spectral{Higgs}$ and $\spectral{CDW}$ continuously.

We attribute the peak in the amplitude spectral functions at $\omega=0$ 
to the freedom to vary the order parameters, i.e., 
diminishing one and increasing the other while keeping $\Delta_\text{tot}$ 
constant so that the system stays in its ground states. 
In the pure SC phase, we find this peak only in $\spectral{CDW}$.
This can be understood as follows: The operator $A_\text{CDW}$ in \eqref{eqn:resolvent_bases} 
creates an exciton and thereby increases $|\Delta_\text{CDW}|$.
This can be achieved without increasing the energy by lowering $\Delta_\text{SC}$.
However, the reversed argument is not true: Since $\Delta_\text{CDW}$ is already 0, 
creating another Cooper pair by virtue of $A_\text{Higgs}$ always increases $\Delta_\text{tot}$ and thus 
requires a minimum energy of $2 \Delta_\text{tot}$.
The analogous argument holds for the pure CDW phase.
If, however, both orders are present, the system can shift between them arbitrarily 
by creating either kind of excitation.
Hence, in this case, we find a peak at $\omega=0$ in both spectral functions.

The spectral function $\spectral{Phase}$ behaves as it does in the pure SC phase. 
It has only a peak at $\omega = 0$ that is associated with the freedom of choice
of the phase of $\Delta_\text{SC}$.


Finally, we investigate the same spectral functions close to the AFM-CDW phase transition
as depicted in Fig.\ \ref{fig:resolvent_overview_AFM}.
We choose $V=1.2t$ for the square lattice and $V=0.8t$ for the simple cubic lattice so that
the phase transition is located at $U=4.8t$ in both cases. 
The panels (a) and (b) show the system close to the phase transition at $U=4.8t \pm 0.05t$.
Again, the SC-related spectral functions are identical in the CDW and the AFM phases
due to the absence of a finite SC order.
Within the continuum, their weight is shifted towards the upper edge.
On the square lattice only, there is a peak above the continuum.
It is likely to be smeared out if lifetime effects were included.
In the AFM phase, both $\spectral{l.AFM}$ and $\spectral{CDW}$ have a peak close to, but yet below, the two-particle continuum. 
The peak of the former is at a lower energy than the peak of the latter.
In the CDW phase, this behavior is reversed, i.e., the peak of $\spectral{CDW}$ lies lower
than the one in $\spectral{l.AFM}$.
Analyzing the power-law behavior of the real part described above, see also
App.\ \ref{sec:fit_greens_functions}, we confirm that these peaks are $\delta$ peaks.
Moving further away from the phase transition, here $U=6.2t$ in panels (c) and $U=3.4t$ in panels (d),
the upper peak merges with the two-particle continuum while the lower one persists.
We emphasize that the spectral function of the transverse magnon $\spectral{t.AFM}$ always 
displays a $\delta'$ peak at $\omega=0$, independent of how far away the parameters are chosen
from the phase transition. This is the expected behavior of a Goldstone boson.


\subsection{SC-CDW transition}

Having discussed the dominant peaks, we address their behavior when 
the system approaches the phase transition between the SC and CDW phase at $V=0$ for $U=-2.5t$.
First, we investigate the identical peak in $\spectral{Phase}$ and $\spectral{Higgs}$, respectively, in the CDW phase.
The evolution of its position $\omega_0$ and its weight $W_0$ as a function of $V$ is depicted
double logarithmically in Fig.\ \ref{fig:sc_in_cdw_behavior}.
For $V \ll t$, the position increases with the square root of $V$. 
This behavior crosses over to a purely linear growth for large $V$.
The gap position $\omega_0$ relative to the gap $\Delta_\text{CDW}$ shows a plateau for large $V$ 
while leaving the square root behavior for small values of $V$ hardly altered.
The peak's weight $W_0$ decreases like $1/\sqrt{V}$ for $V \ll t$,
reaching a shallow minimum for intermediate valus of $V$, and 
eventually saturating at a constant value.

Summarizing, we see that for $V \ll t$ the peak position follows 
$\omega_0 = \alpha \sqrt{V}$ and its weight  $W_0 = \beta / \sqrt{V}$, 
where $\alpha$ and $\beta$ are some constants. Thus, we conclude $W_0 = \beta / (\alpha \omega_0)$. 
Furthermore, we know that the spectral functions are antisymmetric, i.e., we find
\bs
\begin{align}
    \mathcal{A}_\text{Phase} (\omega \ll t) &= W_0 (\delta (\omega - \omega_0) - \delta (\omega + \omega_0)) 
		\\
    &= - \frac{2\beta}{\alpha} \frac{\delta (\omega + \omega_0) - \delta (\omega - \omega_0)}{2\omega_0}.
\end{align} 
\es
Taking the limit $V \to 0$ corresponds to the limit $\omega_0 \to 0$, i.e., leading to a ratio tending to the
derivative
\begin{equation}
    \lim_{h \to 0} \frac{f(x + h) - f(x - h)}{2h} = f'(x).
\end{equation}
Applying this to the spectral peak yields
\begin{align}
    \lim_{\omega_0 \to 0} \mathcal{A}_\text{Phase} (\omega \ll t) = - \frac{2 \beta}{\alpha} \delta'(\omega),
\end{align}
confirming our previous analysis of the phase peak being proportional to  $\delta' (\omega)$.


Next, we analyze the peak in $\spectral{CDW}$ occurring if the system is in the SC phase for $V<0$.
A plot of its position $\omega_0$ and weight $W_0$ is given in Fig.\ \ref{fig:cdw_in_sc_behavior}.
The upper panels (a) and (b) depict the peak position and weight, respectively.
For $|V| \ll t$, the behavior of both of them is the same as for the 
previously discussed peak in $\spectral{SC}$ in the CDW phase.
Increasing $|V|$, the increase of the peak position slows down while the weights start to drop rapidly.
We cannot compute the Green's functions for even larger $|V|$ with $V<0$ because the dynamical matrix $\mM$ 
acquires negative eigenvalues below $V\approx -0.28t$ (square lattice) or $V\approx -0.34t$ (simple cubic lattice). 
We recall that this indicates that the assumed phase, i.e., the SC phase is not stable. 
Therefore, we conclude that for sufficiently negative values of $V$, 
the superconducting phase is not the true ground state as indicated by the red shading in Fig.\ \ref{fig:phase_diagram}.
Specifically, the region of the SC phase on the square lattice 
appears to be even smaller than shown by the $d_{x^2 -y^2}$ data from Ref.\ \cite{Micnas88b}.
Recently, there has been evidence for phase separation for negative values of $V$ 
on the square lattice, at least at finite temperature \cite{Linner23}. 
Thus, we interpret the negative eigenvalues in the dynamical matrix as indicators of phase separation. 
Physically, a strong attractive interaction clearly favors phase separation since the particles prefer to stay close together 
so it is highly plausible that they gather in one region of the sample while leaving the remainder essentially empty.
In 3D, we observe negative eigenvalues in $\mM$ for strongly negative interactions as well.
Thus, we expect that phase separation also occurs on this lattice.


\subsection{AFM-CDW transition}

Last, we investigate how the modes behave as the system passes from the CDW to the AFM phase.
As in the previous subsection, we plot the occurring peaks in $\spectral{CDW}$ and 
$\spectral{l.AFM}$ in the respective phases in Fig.\ \ref{fig:afm_cdw_peaks_overview}.
On the right-hand side of each panel, i.e., for $U > zV$, the system is in the AFM phase (green shading);
on the left-hand side, it is in the CDW phase (orange shading).
Qualitatively, the peak in $\spectral{l.AFM}$ behaves exactly like the peak in 
$\spectral{CDW}$ if the phases are swapped.
We attribute this behavior to the simplicity of our model and do not expect this behavior quantitatively generic.

Panel (a) shows the peak positions which shift essentially linearly upon varying $U$.
It is noteworthy, that in the AFM phase the peak in $\spectral{l.AFM}$ lies energetically
lower than the one in $\spectral{CDW}$.
This is reversed if the system is in the CDW phase.
We conclude that it is energetically more favorable to create an AFM excitation (magnon) than a CDW
exciton if the system is in the AFM phase. In the CDW phase, the creation of the CDW exciton is
energetically cheaper than the creation of a magnon.

Additionally, we plot the peak positions in panel (b) relative to the lower edge of the two-particle continuum.
We see that the lower-lying peak, i.e., the peak corresponding to the phase in which the system is, 
shifts further away from the continuum as the system moves away from the phase boundary.
The other peak, however, shifts closer to the continuum until it dives into it.
At this point, the weight of the merging peak vanishes, see panels (c).
This occurs around $U \approx 4.8t + 1.428t$ ($\spectral{CDW}$) and $U \approx 4.8t - 1.428t$ ($\spectral{l.AFM}$) for the square lattice 
and around $U \approx 4.8t + 0.321t$  ($\spectral{CDW}$) and $U \approx 4.8t - 0.321t$  ($\spectral{l.AFM}$) for the simple cubic lattice.

The weights of the peaks in $\spectral{l.AFM}$ in the AFM phase (the same applies to $\spectral{CDW}$ in the CDW phase) grow as the system moves away from the phase transition.
The same applies to the peaks in $\spectral{CDW}$ in the CDW phase.

In Fig.\ \ref{fig:afm_cdw_peaks_details}, we revisit the data of panels (b) and (c) of the previous figure
on a double logarithmic scale.
We restrict the analysis to $\spectral{l.AFM}$ in the CDW phase as the swapped case of $\spectral{CDW}$ in the 
AFM phase shows the same behavior.
We consider $U$ relative to the value $U_0$ at which the peak in $\spectral{l.AFM}$ ceases to exist.
This occurs at $U_0 = 3.372t$ on the square lattice and at $U_0 = 4.479t$ on the simple cubic lattice.
The peak positions $\omega_0$ in panels (a) behave quadratically as they merge with the 
two-particle continuum, i.e., 
$\omega_0 \propto (U-U_0)^2$ while the weights $W_0$ in panel (b) decrease linearly, i.e., 
$W_0 \propto (U-U_0)$. This behavior is generic for the peaks of bound states merging with
their continua at vanishing binding energy \cite{uhrig96b,uhrig96be,zhito13}.


\section{Conclusion}\label{sec:conclusion}


The objective of this article was to study collective excitations in competing phases.
We wanted to understand how collective excitations behave in the vicinity of phase transitions
and how they signal these transitions, for example by softening or by losing spectral weight.
To this end, we developed a general and versatile formalism based on iterated equations of motion.
We could show that the formalism reproduces all rigorously known properties such as
the existence of Goldstone bosons in case of broken continuous symmetries and the absence thereof
if the broken symmetry is discrete.

We conducted an analysis of the mean-field phase diagram of the extended Hubbard model in 
two (2D) and three dimensions (3D). The results in both dimensions are qualitatively very similar.
Specifically, we studied the competing phases 
of $s$-wave superconducting order (SC), alternating charge density wave order
(CDW), and N\'eel antiferromagnetism (AFM).
Furthermore, we were able to identify regions in which our analysis is incomplete
in the sense that the assumed phases turn out to be unstable, but it is not obvious which
phases are the stable ones. Since we observed this phenomenon for strongly attractive interactions
we conjecture that the instability indicates the occurrence of phase separation in 2D and 3D. 
For the square lattice, recent findings at finite temperature also pointed in this direction \cite{Linner23}.

We developed a method to obtain Fourier-transformed Green's functions
based on iterated equations of motion and analyzed the most relevant
of them. We identified the signatures of collective excitations in the spectral functions 
$\spectral{Higgs}$, $\spectral{Phase}$, $\spectral{CDW}$, $\spectral{l.AFM}$, and $\spectral{t.AFM}$ 
which are designed to excite the Higgs mode, the phase mode, a staggered density mode, a longitudinal, and a transversal magnon, respectively. 
The focus was the analysis of their behavior as the system approaches and crosses phase transitions.
In particular, we identified the peak and the singularity occurring in the first two 
spectral functions related to superconductivity
to be the well-known amplitude and phase modes in neutral superconductors 
and thereby provided a technique to obtain them starting from a microscopic description.
Outside an SC phase, both spectral functions are identical while in an SC phase,
the phase mode is massless while the Higgs mode is at the lower edge of the continuum.
So, the phase mode manifests as the derivative of the $\delta$ function at zero energy
complying with the asymmetry of the spectral function.

We observed the behavior of the peak in the SC-related spectral functions corresponding to a cooperon in various phases. 
It shifts towards zero energy and acquires more and more weight as the system moves from the CDW phase toward the SC phase. 
The peak's relative position to the two-particle continuum and its weight saturate 
as the system moves away from the phase transition.
In the SC phase close to the CDW phase, the peak in $\spectral{CDW}$ 
is located at a finite energy and and has finite weight. 
It must represent a bound state because it is located below the two-particle continuum. 
This state is induced by the application of an electron creation and an electron annihilation without effect of its spin content so that we interpret it as an $S=0$ exciton.
It behaves similarly to the SC-related peak in non-SC phases.

In the vicinity of the AFM-CDW phase transition, the spectral densities
$\spectral{CDW}$ and $\spectral{l.AFM}$ feature a peak. 
The former is the aforementioned exciton while the latter is a longitudinal magnon. 
Of course, this can also be seen as an exciton being a bound
state of an electron and a hole, yet with spin content $S=1$ and $S^z=0$. 
The longitudinal magnon lies at lower energy in the AFM phase than the CDW exciton. 
In the CDW phases, the two peaks swap their relative energy positions. 
Moving away from the phase transition, the higher-lying peak merges with the two-particle continuum. 
The peak position relative to the two-particle continuum approaches $0$ 
quadratically in $\Delta U$ while its weight approaches $0$ linearly with $\Delta U$ where
$\Delta U$ measures the distance to the interaction where the merging takes place.

Furthermore, we found the exciton peak in $\spectral{CDW}$ also at large values of $V$ in the CDW phase, 
evolving smoothly from the CDW-AFM phase transition. 
We assume that this peak is always located outside of the two-particle continuum, even for small values of $V$.
In this case, however, it can lie very close to the continuum, so that we cannot resolve it due to the limited accuracy of our numerics.
So it seems to appear as a singularity at $2\Delta$ for small values of $V$.

In the AFM phase, the spectral functions $\spectral{l.AMF}$ and $\spectral{t.AFM}$ are distinct.
While the longitudinal magnon continues to manifest as a peak at relatively high energies
the transversal magnon appears as a peak at zero energy within our numerical accuracy. Due to
the asymmetry of the spectral, its peak is the derivative of a $\delta$ function.
It is an advantageous feature of the adopted approach that the Goldstone theorem is 
automatically fulfilled. Both, phase mode and transversal magnon are located at zero energy
if the corresponding symmetry is broken, i.e., the U(1) symmetry and the O(3) symmetry,
respectively.


Future directions of research comprise incorporating the investigation of possible phase-separated states. 
This would enable us to delve deeper into the negative $V$ regime 
of the phase diagrams and to analyze its collective excitations. Anticipating the occurrence
of regions of high and low density of electrons, a necessary prerequisite is to investigate doped systems.

Another intriguing extension is to extend the study from the $\Gamma$ point to the full Brillouin zone.
This implies studying the same phases with operators at finite wave vectors beyond the nesting vector. 
Conceptually, this extension is straightforward. But it spoils the efficient approach
to discretize the $\gamma$ space since the dependence on two wave vectors will no longer
allow us to reduce all wave vector dependence to a single dependence on $\gamma$.

Additionally, one can fathom including a coupling of the electronic system to the electromagnetic fields
so that we no longer study a neutral, but a charged superconducting system. One would expect
to capture the important shift of the phase mode to finite energies close to the plasma frequency 
as explained by the Anderson-Higgs mechanism.

Lastly, one could deal with multi-band systems with even richer phase
diagrams and corresponding collective excitations.

\begin{acknowledgments} 
	This research was partially funded by the MERCUR Kooperation
		in project KO-2021-0027. We acknowledge very helpful discussions with
		Anna B\"ohmer, J\"org B\"unemann, Ilya Eremin, G\"otz Seibold, Joachim Stolze, and Zhe Wang
		as well as with all members of our research group.
\end{acknowledgments}

\appendix
\section{Semipositivity of the dynamical matrix}
\label{sec:positive_M}

We argue that $\mathcal{M}$ is positive semidefinite if the system is in thermal equilibrium.
For any $\vec{x} \in \mathbb{C}^N$, we consider
\bs
\begin{align}
    \vec{x}^\dagger \mM \vec{x} &= \sum_{ij} x_i^* x_j \mM_{ij} 
		\\
        &= \Big\langle \Big[ \sum_i x_i^* A_i^\dagger, \Big[ H, \sum_j x_j A_j \Big] \Big]  \Big\rangle 
				\\
        &= \langle [B^\dagger, [H, B]] \rangle,\quad B \coloneqq  \sum_j x_j A_j.
\end{align}
\es
For $\mM$ to be positive semidefinite the expression above must not be negative.
Exactly this property has been already proven in Refs.\ \cite{mermin66,Dyson1978}
as the non-negativity of the double commutator.

\section{Fits to the Green's functions}
\label{sec:fit_greens_functions}

\begin{figure*}[htb]
    \centering
    \includegraphics[width=0.98\textwidth]{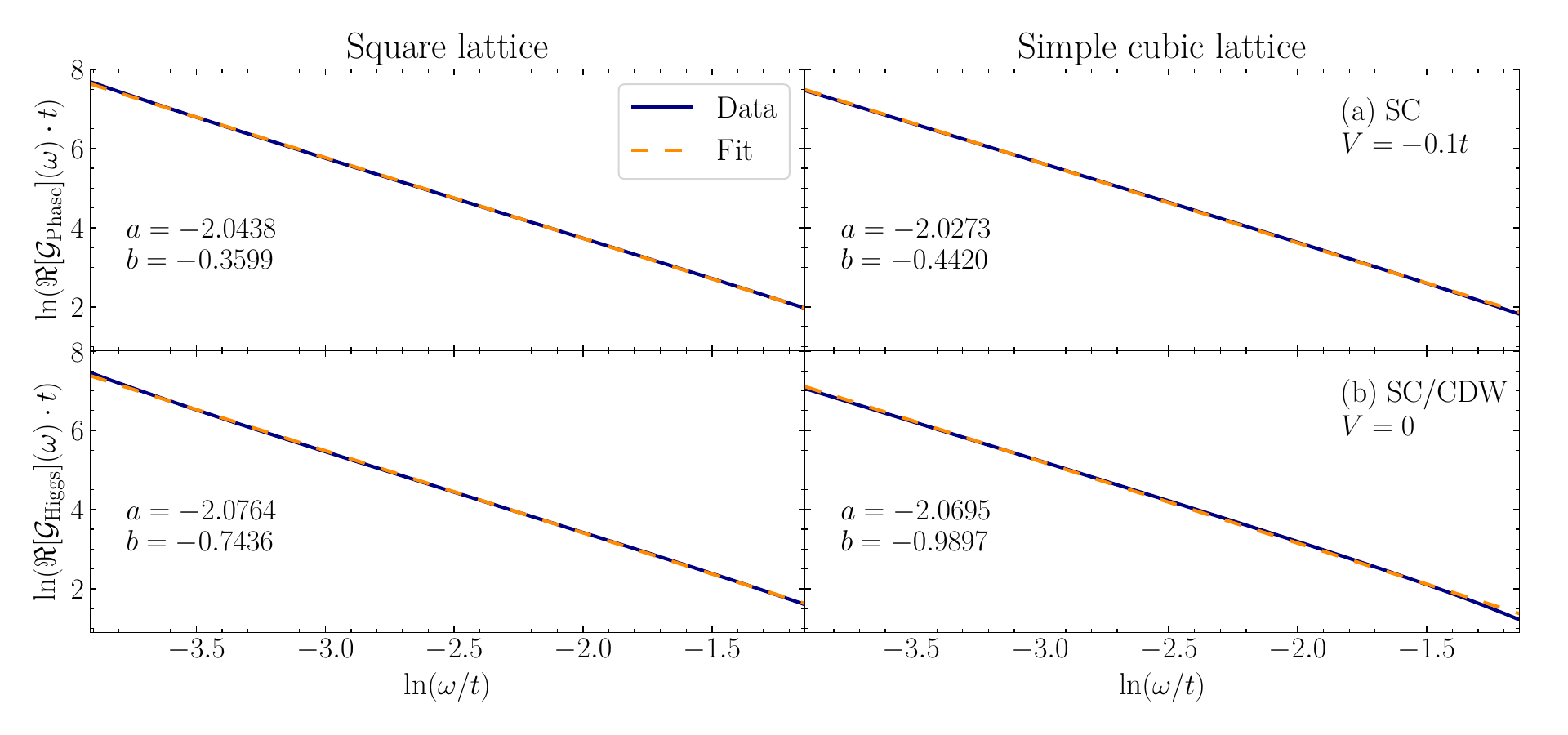}
    \caption{The upper panel shows a log-log plot of the real part of $\spectral{Phase}$ 
		in the SC phase at $U=-2.5t$ and $V=-0.1t$ as well as a linear fit to it.
    The lower panel shows the same plot for $\spectral{Higgs}$ in the coexistence phase at $V=0$. 
		The columns show the results for the square and the simple cubic lattice, respectively.
    The functions behave as $1/\omega^2$ indicating that the peaks in the spectral functions are derivatives 
		of a $\delta$ distribution.}
    \label{fig:zero_peaks}
\end{figure*}

Here, we discuss fits to the real parts of various Green's functions in the different phases.
All fits are linear of the type $y = ax + b$ in double logarithmic plots, i.e., $a$ describes the exponent of
the power-law behavior of the functions.
An examplary plot of the real part of $\greens{Phase}$ 
in the SC phase is shown in panel (a) of Fig.\ \ref{fig:zero_peaks}.
It behaves like $1/\omega^2$ indicating that the peak 
at $\omega=0$ is the derivative of a $\delta$ distribution.
The same kind of result is obtained for the transversal magnon in $\spectral{t.AFM}$ in the AFM phase, 
as well as in $\spectral{Higgs}$ and $\spectral{CDW}$ in the coexistence phase.

The analogous analysis of the divergence at $\omega = 2\Delta$ 
found in $\spectral{Higgs}$ and $\spectral{CDW}$ at $V=0$
reveals the power law  $1/\sqrt{\omega}$.

Lastly, all other peaks below the two-particle continuum behave identically.
The real part of them behaves like $1/\omega$ in close vicinity to the peak position.
This indicates that the peaks are $\delta$ distributions.
The peaks of $\spectral{l.AFM}$ and $\spectral{CDW}$ close to the phase transition are identical 
when swapped, i.e., $\spectral{l.AFM}$ in the CDW phase is the same as $\spectral{CDW}$ in the AFM phase.
However, we do not believe that this identity is quantitatively generic.

\section{Numerical treatment of the matrix elements in $\gamma$-space}
\label{sec:numerical_ieom}

We want to use operators such as \eqref{eqn:ieom_basis_operator} to span the considered operator set.
Numerically, however, it is impossible to deal with matrices of infinite dimensions so a discretization is necessary.
Additionally, we cannot use $\delta$ distributions as matrix elements. 
Therefore, we also discretize the $\delta$ distribution as follows.
A mesh of equidistant sampling points is chosen for $\gamma\in[-1,1]$.
They are midpoints of intervals of length $\Delta \gamma$.
Then, we define an approximate $\delta$ function by
\begin{equation}
    h(\gamma) \coloneqq \begin{cases}
        \frac{1}{\Delta \gamma} & |\gamma| < \frac{\Delta \gamma}{2} \\ 0 & \text{otherwise}
    \end{cases}.
\end{equation}
Numerical integration amounts up to the finite sum over the sampling points and for the approximate $\delta$ function in particular reads
\begin{equation}
    \int_{-1}^1 f(\tilde{\gamma}) h(\tilde{\gamma} - \gamma_j) \mathrm{d} \tilde{\gamma} \approx 
		\sum_i f(\gamma_i) h(\gamma_i - \gamma_j) \Delta \gamma = f(\gamma_j),
\end{equation}
showing that $h(\gamma)$ mimics the continuous $\delta$ distribution on the discrete mesh.
Note that the limit $\Delta \gamma \to 0$ reproduces the continuous case with the $\delta$ distribution.

We compute the matrix elements as
\begin{equation}
    \label{eqn:numerical_example}
    (A_i | A_j) = \frac{1}{N} \sum_{\vk \vl} h(\gamma_i - \widehat{\gamma}(\vk)) h(\gamma_j - 
		\widehat{\gamma}(\vl)) (A_{\vk} | A_{\vl}).
\end{equation}
Here, all expressions represent finite numbers so that they are suitable for
numerical treatment. This procedure can also be applied to the matrix elements $(A_i | [H, A_j])$ of the dynamical matrix.

Additionally, we stress that expressions such as $(A_{\vk} | A_{\vl})$ or
$(A_{\vk} | [H, A_{\vl}])$ can be reduced to sums proportional to $\gamma_i$ or $\gamma_j$, respectively.
Therefore, it is not necessary to implement any sort of computation in reciprocal space.
As an example, let us consider $A_{\vk} = f_{\vk}^\dagger$ and 
\begin{equation}
    ( f_{\vk}^\dagger | f_{\vl}^\dagger ) = \delta_{\vk, \vl}\, \langle 1 - n_{\vk \uparrow} - n_{\vk \downarrow} \rangle = \alpha(\vk) \delta_{\vk, \vl}  
\end{equation}
with $\alpha(\vk)\coloneqq 1 - \langle n_{\vk \uparrow} + n_{\vk \downarrow} \rangle$. 
Inserting this into Eq.\ \eqref{eqn:numerical_example} yields
\begin{equation}
    (A_i | A_j) = \frac{1}{N} \sum_{\vk} \alpha(\vk) h(\gamma_i 
		- \widehat{\gamma}(\vk)) h(\gamma_j - \widehat{\gamma}(\vk)).
\end{equation}
As stated in the main text, see Eq.\ \eqref{eqn:equal_expecs}, $\alpha(\vk)$ depends only on $\widehat{\gamma}(\vk)$, 
hence we can pass from the wave vector integration to a $\gamma$ integration and eventually to a $\gamma$ sum
\bs
\begin{align}
    (A_i | A_j) &= \int \rho(\gamma) h(\gamma_i - \gamma) h(\gamma_j - \gamma) \alpha(\gamma) \dgamma 
      \\
    &\approx \rho(\gamma_i) h(\gamma_j - \gamma_i) \alpha(\gamma_i) 
			\\
    &= \frac{\delta_{ij}}{\Delta \gamma} \rho(\gamma_i) \alpha(\gamma_i).
\end{align}
\es
Since $\Delta \gamma$ is a positive constant factor, it does not impact any of our matrix operations, i.e., we can shift it to the front of the expressions $\mM \to \tilde{m} (\Delta \gamma)^2$ 
and $\mN \to \tilde{n} (\Delta \gamma)^2$.
This allows us to rewrite \eqref{eqn:green_derivation} as
\begin{equation}
    G_{AB^\dagger}(z) = - (\Delta \gamma)^2 \vec{b}^\dagger (0) \left[ \tilde{n} \frac{1}{\tilde{m} + z \tilde{n}} \tilde{n} \right] \vec{a}(0)
\end{equation}
for the general operators $A$ and $B$ and the initial conditions
$A=\sum a_i A_i$ and $B=\sum b_i A_i$.

In order to provide an explicit example, 
let us compute $\greens{Higgs}$ from \eqref{eqn:resolvent_bases}.
We order our operator set such that the operators $f_\gamma + f_\gamma^{\dagger}$ are the first ones
in the set. Then, the initial conditions read 
\begin{equation}
    a_i = b_i = \begin{cases}
        1 & 1\le i \le N_\gamma \\
        0 & \text{otherwise}
    \end{cases},
\end{equation}
where $N_\gamma$ is the number of sampling points.
For the Green's function, this implies
\bs
\begin{align}
    \green{Higgs} (z) &= -(\Delta \gamma)^2 \sum_{i=1}^{N_\gamma} \sum_{j=1}^{N_\gamma} \left[ \tilde{n} \frac{1}{\tilde{m} + z \tilde{n}} \tilde{n} \right]_{ij} 
		\\
        &\approx - \int \dgamma \int \dgamma' \left[ \tilde{n} 
				\frac{1}{\tilde{m} + z \tilde{n}} \tilde{n} \right](\gamma, \gamma').
\end{align}
\es
The first line is an approximation of the integral in the second line, as each matrix element corresponds to a specific value for $\gamma$.

\section{Influence of the discretization}
\label{sec:finite_size}

\begin{figure}[htb]
    \centering
    \includegraphics[width=0.48\textwidth]{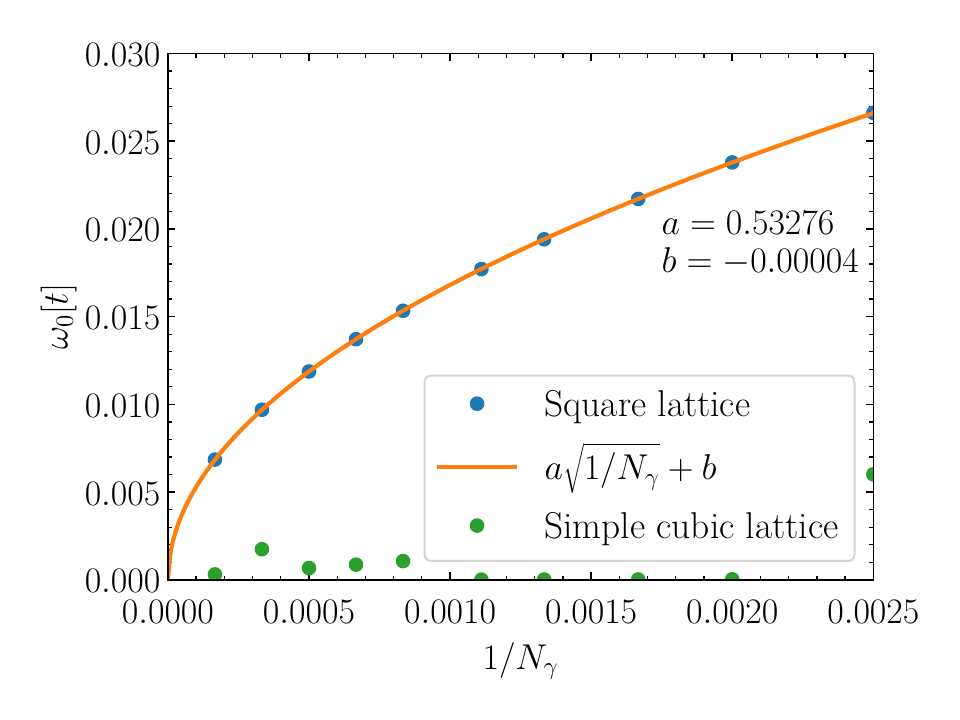}
    \caption{Effect of the number of  sampling points $N_\gamma$ 
		on the absolute position of the peak in $\spectral{Phase}$ for $U=-2.5t$ and $V=-0.1t$.
    On the simple cubic lattice, the peak already is located at $\omega=0$ within numerical accuracy.
    On the square lattice, its position approaches $0$ following $\sqrt{1/N_\gamma}$.}
    \label{fig:finite_size}
\end{figure}

Most of the computed data barely depend on the number of sampling points $N_\gamma$.
However, specific features at very small energies can be difficult to  resolve properly, e.g., 
the peak in $\spectral{Phase}$ in the SC phase.
We illustrate this effect in Fig.\ \ref{fig:finite_size}.
Neglecting minor numerical scattering, the phase peak is properly located at $\omega=0$ for the simple cubic lattice.
On the square lattice, however, it follows a $\sqrt{1/N_\gamma}$ behavior, i.e., also approaching $0$,
but only in the limit $N_\gamma \to \infty$.
This behavior can be observed for the other peaks at $\omega=0$ as well.
How strong this effect is, i.e., how large the deviation from $0$ is depends on the spectral function
as well as on the other parameters of the system, in particular its dimension.

\bibliography{sn-bibliography}
		
\end{document}